\documentclass[review,10pt]{elsarticle}
\usepackage{amssymb}
\usepackage{geometry}
\journal{:}
\usepackage{ae}
\usepackage[T1]{fontenc}
\usepackage[ansinew]{inputenc}
\usepackage{amsmath}
\usepackage{color}
\usepackage[colorlinks]{hyperref}
\usepackage{graphicx}% Include figure files
\usepackage{dcolumn}% Align table columns on decimal point
\usepackage{bm}% bold math
\usepackage{hyperref}% add hypertext capabilities
\usepackage[mathlines]{lineno}% Enable numbering of text and display math
\usepackage{xspace}
\usepackage{algorithm}
\usepackage{algorithmic}
\usepackage{subfigure}
\usepackage{amsfonts}
\usepackage{enumerate}
\usepackage{framed}
\usepackage{color}
\usepackage{amsthm}
\input{epsf}
\newcommand{\figref}[1]{{Figure~\ref{#1}}}

\newcommand{\secref}[1]{{Section~\ref{#1}}}
\renewcommand{\eqref}[1]{{Eq.~(\ref{#1})}}

\newcommand{\sech}{\mbox{\textrm{\ensuremath{\,}sech}}}

\begin{document}
\begin{frontmatter}
\title{\textbf{Nonlinear dynamics of damped DNA systems with long-range interactions}}
\author[biophys,cetic]{\textbf{J. Brizar Okaly}\corref{cor1}} \ead{okalyjoseph@yahoo.fr} \cortext[cor1]{Corresponding author}
\author[biophys,cetic]{\textbf{Alain Mvogo}} \ead{mvogal\_2009@yahoo.fr}
\author[meca,cetic]{\textbf{R. Laure Woulaché}} \ead{rwoulach@yahoo.com}
\author[meca,cetic]{\textbf{T. Crépin Kofané}} \ead{tckofane@yahoo.com}
\address[biophys]{Laboratory of Biophysics, Department of Physics,
Faculty of Science, University of Yaounde I, P.O. Box 812, Yaounde,
Cameroon}
\address[meca]{Laboratory of Mechanics,
Department of Physics, Faculty of Science, University of Yaounde I,
P.O. Box 812, Yaounde, Cameroon}
\address[cetic]{African Center of Excellence in Information and Communication
Technologies, University of Yaounde I, P.O. Box 812, Yaounde,
Cameroon.}
\date{\today}
\begin{abstract}
We investigate the nonlinear dynamics of a damped Peyrard-Bishop DNA model taking into account long-range interactions with distance dependence $|l|^{-s}$ on the elastic coupling constant between different DNA base pairs. Considering both Stokes and long-range hydrodynamical damping forces, we use the discrete difference operator technique  and show in the short wavelength modes that the lattice equation can be governed by the complex
Ginzburg-Landau equation. We found analytically that the technique leads to the correct expression for the breather soliton parameters. We found that the viscosity makes the amplitude of the breather to damp out. We compare the approximate analytic results with numerical simulations for the value $s=3$ (dipole-dipole interactions).
\begin{keyword}
DNA, long-range interactions, damping forces,  breather soliton.
\end{keyword}
\end{abstract}
\end{frontmatter}
%
%\tableofcontents
%
\section{\label{introduction} Introduction}
The DNA molecule is known to be very important and essential in the protection, transport and transmission of the genetic code. Many theoretical models have been proposed to describe the nonlinear dynamics of DNA. The first nonlinear DNA model was suggested by Englander \emph{et al.} \cite{Englander}. Thereafter, simplified models were proposed to describe the angular distortion of DNA \cite{Yomosa1,Homma1,Homma2} and the micromanipulation experiments
were investigated, showing the great importance of radial displacements of bases during the processes of replication and transcription \cite{PB,Dauxois1,Dauxois2}. The Peyrard-Bishop (PB) DNA model \cite{PB}, which has been successfully  used to analyze experiments on short DNA sequences \cite{Campa} has gained a popularity in that direction.

DNA double helix spontaneously denatures locally and the breathing mode occurs when it is locally excited with large amplitude \cite{dau}. The amplitude of this excitation depends on the surrounding environment-DNA interactions and the inner mechanism of the DNA molecule depending on the relative motion of particles. It is therefore of physical importance to take into account these two effects in the nonlinear dynamics of DNA molecule. Some authors
emphasized the influence of the viscosity on the dynamical properties of the DNA molecule \cite{zdr2,zdr22}. In particular, it
is shown that the viscosity of the medium damps out the amplitude of the nonlinear wave propagating through the molecule \cite{zdr2,zdr22}. In the above studies, the analysis deal with the PB DNA model with nearest neighbor interactions between base pairs. The long-range interactions (LRI) play a crucial role in molecular systems \cite{Calla,Sha,rau}. The importance of LRI in DNA molecule is due to the presence of phosphate groups along the strands \cite{Calla}. The LRI therefore allow to take into account the screening of the interactions or an indirect coupling between base pairs (e.g. via water filaments) \cite{Sha}. Rau and Parsegian, in their experimental studies in the direct measurements of the intermolecular forces between counterion-condensed DNA double helices, have shown the importance of long-range attractive hydrogen forces and emphasize that many forces could be responsible for LRI in the DNA molecule \cite{rau}. In fact, the charged groups in the molecular chains and DNA molecules interact through long-range
dipole-dipole interactions. Along the same line, it has been shown that the study of the power-law LRI in nonlinear lattice models is very relevant mainly in molecular chains and DNA molecules where Coulomb and dipole-dipole interactions are of a great physical importance \cite{ish,gai1,gai2}. Recent works by Mvogo \emph{et al.} \cite{mvo1,mvo2} also indicated qualitative effects of the power-law LRI in molecular systems.

To the best of our knowledge, no work has been reported on the study of DNA dynamics taking into account both damping and LRI effects. Our aim in this paper is to study the DNA dynamics taking into account LRI between different DNA base pairs and both Stokes and long-range hydrodynamical damping effects. To address this issue, our analytical study has been inspired by the one recently developed by Miloshevich \emph{et al.} \cite{mil} to investigate traveling solitons in long-range oscillator chains. Due to the non analytical properties of the dispersion relation, Miloshevich \emph{et al.} \cite{mil} have shown that  the discrete difference operator (DDO) technique is more appropriate to study physical systems with LRI. In this paper, we use the DDO and show that the DNA models with LRI can be reduced to a specific form of the complex Ginzburg-Landau (CGL) equation, where the dispersion coefficient is complex and the nonlinearity coefficient is real. A similar equation has been obtained by Zdravkov\'{i}c \emph{et al.} \cite{zdr2, zdr22} while studying the effect of viscosity on the dynamics of the Peyrard-Bishop-Dauxois (PBD) DNA model in the absence of hydrodynamical damping and LRI forces. The investigators state that the CGL equation cannot be solved analytically like a nonlinear Schrodinger (NLS) equation \cite{zdr2, zdr22}. In this paper, following the work by Pereira and Stenflo \cite{Stenflo}, we analytically solve the CGL equation.

The paper is organized as follows. In \secref{model}, we propose the Hamiltonian model and derive the discrete equations of motion for the in-of-phase and out-of-phase motions. In \secref{method}, we use the DDO and show that the out-of-phase dynamical equation can be reduced to the CGL equation. The envelope soliton solution of this equation is reported and the breather solution of the discrete equation of motion for the out-of-phase motion is derived. In \secref{solution}, we perform the numerical simulations with emphasis on  the effects of the LRI and damping forces. \secref{conclusion} concludes the work.

%%%%%%%%%%%%%%%%%%%%%%%%%%%%%%%%%%%%%%%%%%%%%%%%%%%%%%%%%%%%%%%%%%%%%%%%%%%%%%%%%%%%%%%%%%%
\section{\label{model} Model and equations of motion}

We consider the PB model \cite{PB} for DNA denaturation where the degrees of freedom $x_n$ and $y_n$ associated to each base pair correspond to the displacements of the bases from their equilibrium positions along the direction of the hydrogen bonds that connect the two bases in a pair. A LRI coupling between the base pairs due to the presence of phosphate group along the DNA strands is assumed so that the Hamiltonian for the model is given by

\begin{equation}\label{eq1}
\begin{split}
H =& \sum\limits_n^N\Big\{\frac{1}{2}m(\dot x_n^2 + \dot y_n^2)+
\frac{1}{2}\sum\limits_{l=1} J_l[(x_n - x_{n - l})^2
+(y_n - y_{n - l})^2] + V(x_n,y_n)\Big\},
\end{split}
\end{equation}
where $m$ is the average mass of the nucleotides and $N$ represents the number of the base pairs of the DNA molecule. The interactions between hydrogen bonds in a pair is modeled by the Morse potential
$V(x_n,y_n)$ given by
\begin{equation}\label{eq2}
V(x_n,y_n)=D\left[e^{- a(x_n-y_n)} - 1 \right]^2,
\end{equation}
where $D$ is the depth of the Morse potential well, which may depend on the type of base pair and $a$ is the width of the well. The quantity
\begin{equation}\label{eq3}
J_l=J|l|^{-s},
\end{equation}
is the power-law dependence of the elastic coupling constant, where
$s$ and $|l|$ are the LRI parameter and the normalized distance
between base pairs, respectively. In practice, to keep the spatial
homogeneity in a finite DNA system with periodic boundary
conditions, usually the LRI is limited in each direction to
$\frac{1}{2}(N-1)$, if $N$ is odd, or $\frac{1}{2}(N-2)$,  if $N$ is
even, and $1\leq |l|\leq \frac{1}{2}(N-1)$ \cite{fla}. The parameter
$s$ can be used to model Coulomb interactions between charged
particles of a chain ($s=1$), dipole-dipole interactions ($s=3$).
Below $s=1$ the energy diverges and above $s=3$ the system becomes
short-range. In this paper, we use the case $s=3$, where multiple
solutions exist \cite{Gai3}.

The values of parameters used to perform our analysis are those from
the dynamical and denaturation properties of DNA. They are
\cite{pey}: $m=300$ amu, $J=0.06$ eV/\AA$^{2}$, $D=0.03$ $eV$ and
$a=4.5$ \AA$^{-1}$. Our system of units (amu, \AA, eV) defines a
time unit ($t.u.$) equal to $1.018\times 10^{-14}$ s.

To describe the motions of the two strands, we introduce the new
variables $u_n$ and $v_n$ such that
\begin{equation}\label{eq4}
u_n = \frac{x_n+y_n}{\sqrt 2}\;\ \mathrm{and} \;\ v_n =
\frac{x_n-y_n}{\sqrt2},
\end{equation}
where $u_n$ and $v_n$  represent the in-phase and the
out-of-phase motions. Taking into account \eqref{eq2} and
\eqref{eq4}, the Hamiltonian of the system can be rewritten as
\begin{equation}\label{eq5}
\begin{split}
H =&\sum\limits_n^N\Big\{\frac{1}{2}m\dot u_n^2 +
\frac{1}{2}\sum\limits_{l=1} J_l(u_n- u_{n-l})^2\Big\}+\sum\limits_n^N\Big\{\frac{1}{2}m\dot v_n^2+\frac{1}{2}
\sum\limits_{l=1} J_l(v_n-v_{n - l})^2+ D\left(e^{-a\sqrt2v_n}-1\right)^2\Big\}.
\end{split}
\end{equation}
The equations of motions of the system then read
\begin{equation}\label{eq6}
m\ddot u_n=\sum\limits_{l=1} J_l\left(u_{n+l}-2u_n+u_{n-l}
\right),
\end{equation}
\begin{equation}\label{eq7}
\begin{split}
m {\ddot v_n} =&\sum\limits_{l=1} J_l\left(v_{n+l}- 2{v_n}+v_{n-l}\right)
+2\sqrt2aDe^{-a\sqrt2v_n}\left(e^{-a\sqrt2v_n}-1\right).
\end{split}
\end{equation}
For a more realistic study of dynamical properties of DNA, one must
take into account its environment. In the present work, we take into
account the Stokes ($F^{st}$) and the long-range hydrodynamical
($F^{hy}$) damping forces in the equations of motion of the system.
These forces account respectively for DNA molecules in a viscous
environment and their inner mechanism. The Stokes damping forces is
given by
\begin{equation}\label{eq8}
F_n^{st}=-m \gamma ^{st}\dot q_n,
\end{equation}
where $\gamma^{st}$ is the Stokes damping constant. The coordinate
$q_n$ can be replaced by $u_n$ or $v_n$. In previous works in
discrete lattices, investigators assume the hydrodynamical damping
force in the nearest neighbor interactions \cite{are,bru,isa, pey3}.
In this work, the DNA molecule is considered as a collection of
nucleotides linked to the neighbors of the same strand by spring.
Each of them is assumed to be point masses of mass $m$. Thus, the
displacement of one base pair causes a more or less significant
displacement of the other base pairs of the chain according to
whether they are closed or distanced from the initial base pair. This
displacement gives rise to the hydrodynamical viscous forces which
influence the motion of the initial nucleotide. Since our work
focuses on LRI between base pairs, we have introduced the LRI in the
hydrodynamic dissipation $F^{hy}$ in order to takes into account the
contribution of all base pairs of the chain so that,
\begin{equation}\label{eq9}
F_n^{hy}=m\sum\limits_{l=1}\gamma_l\left(\dot q_{n-l}-2\dot q_n+\dot
q_{n+l}\right),
\end{equation}
where $\gamma_l = \gamma^{hy}|l|^{-s'}$ and $\gamma^{hy}$ is the
hydrodynamical damping coupling constant.
Taking into account $F_n^{st}$ and $F_n^{hy}$ as defined above, we
obtain the following equations of motion:
\begin{equation}\label{eq11}
\begin{split}
\ddot u_n =&\sum\limits_{l=1}\frac{J_l}{m} (u_{n+l}-
2u_n+u_{n-l})-\gamma ^{St}\dot u_n +\sum\limits_{l=1}\gamma
_l\left(\dot u_{n+l}-2\dot u_n+\dot u_{n-l}\right)
\end{split}
\end{equation}
\begin{equation}\label{eq12}
\begin{split}
\ddot v_n = &\sum\limits_{l=1}{\frac{J_l}{m}}(v_{n+l}
-2v_n + v_{n - l})+\frac{2\sqrt 2 aD}{m}e^{ - a\sqrt 2 v_n}(e^{-a\sqrt2
v_n}- 1)-\gamma ^{St}\dot v_n +\sum\limits_{l = 1}
\gamma_l\left(\dot v_{n +l} - 2\dot v_n + \dot v_{n - l}  \right).
\end{split}
\end{equation}
The solution $u_n(t)$ of \eqref{eq11} is an ordinary solution of a
damped linear schr\"{o}dinger equation and represents a plane wave
in a viscous medium in the presence of LRI. In what follows, the
system will be considered heavily damped. Our investigations will be
limited to the analysis of the dynamical behavior of the stretching
motion of each base pair represented by the solution of
\eqref{eq12}, in the presence of LRI and the \textquotedblleft big
viscosity \textquotedblright\; \cite{zdr2, zdr22}.

%%% ----------------------------------------------------------------------
\section{\label{method} Discrete difference operator  technique}
\begin{figure}[H]
\centering
\includegraphics[width=3in]{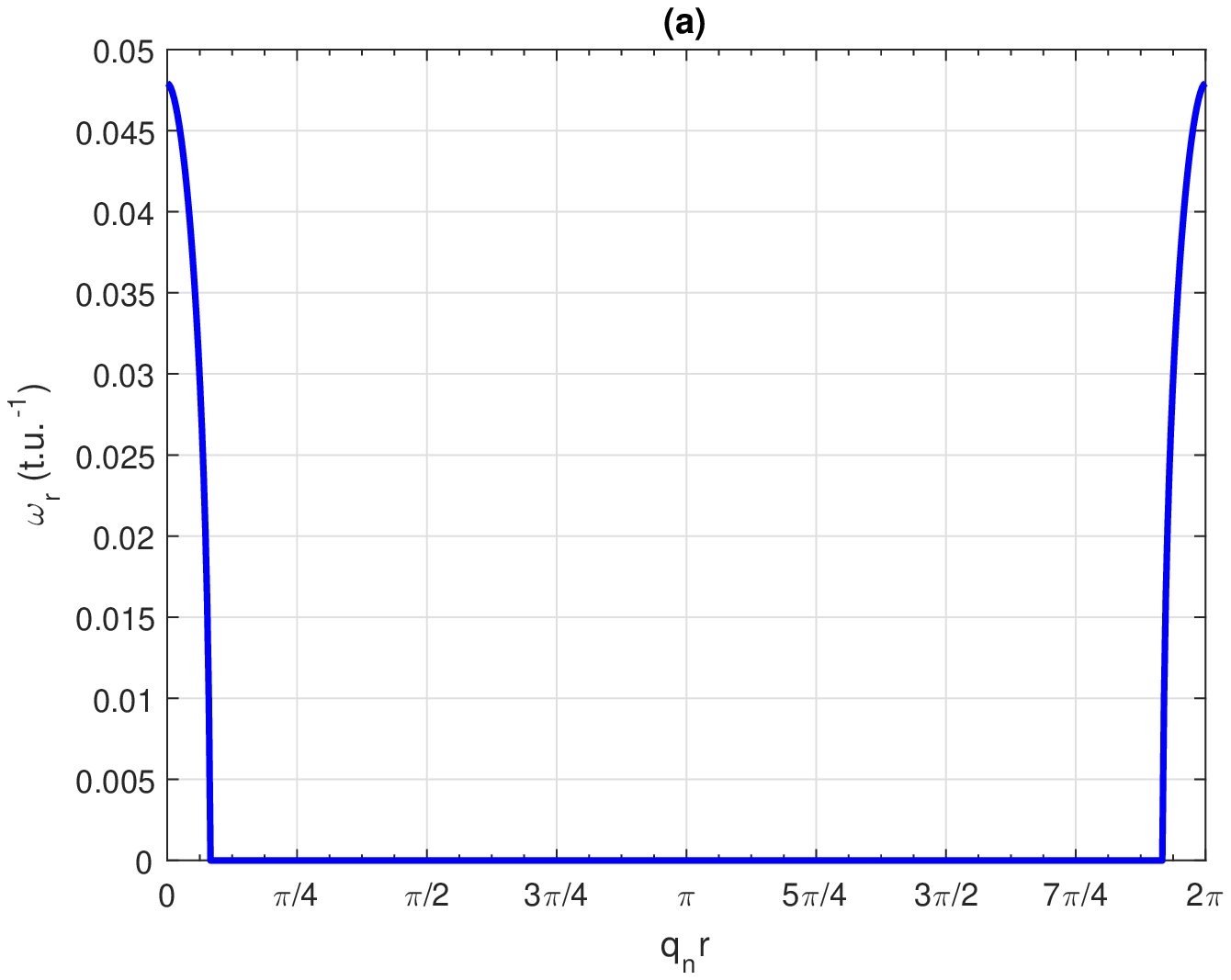}
\includegraphics[width=3in]{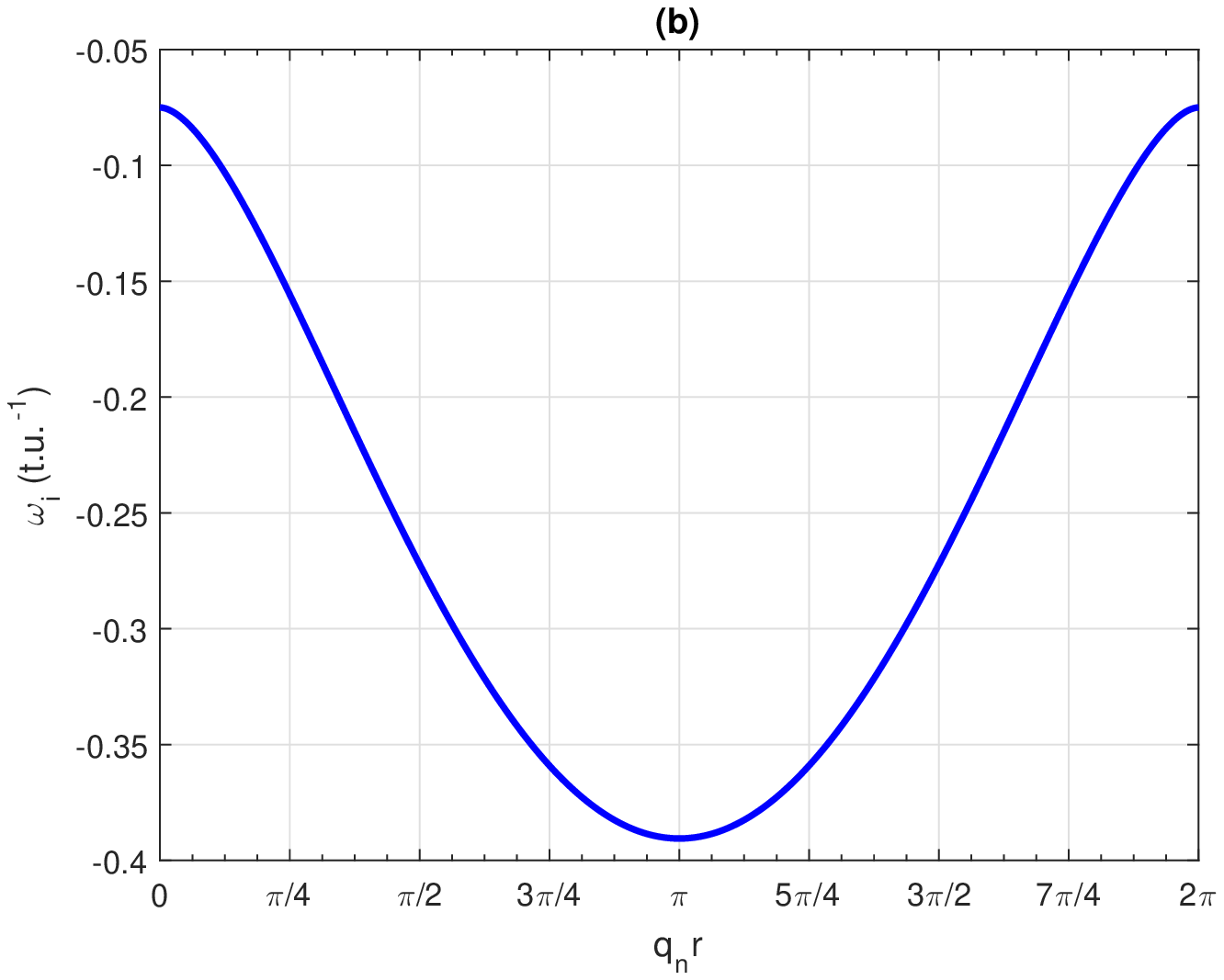}
\includegraphics[width=3in]{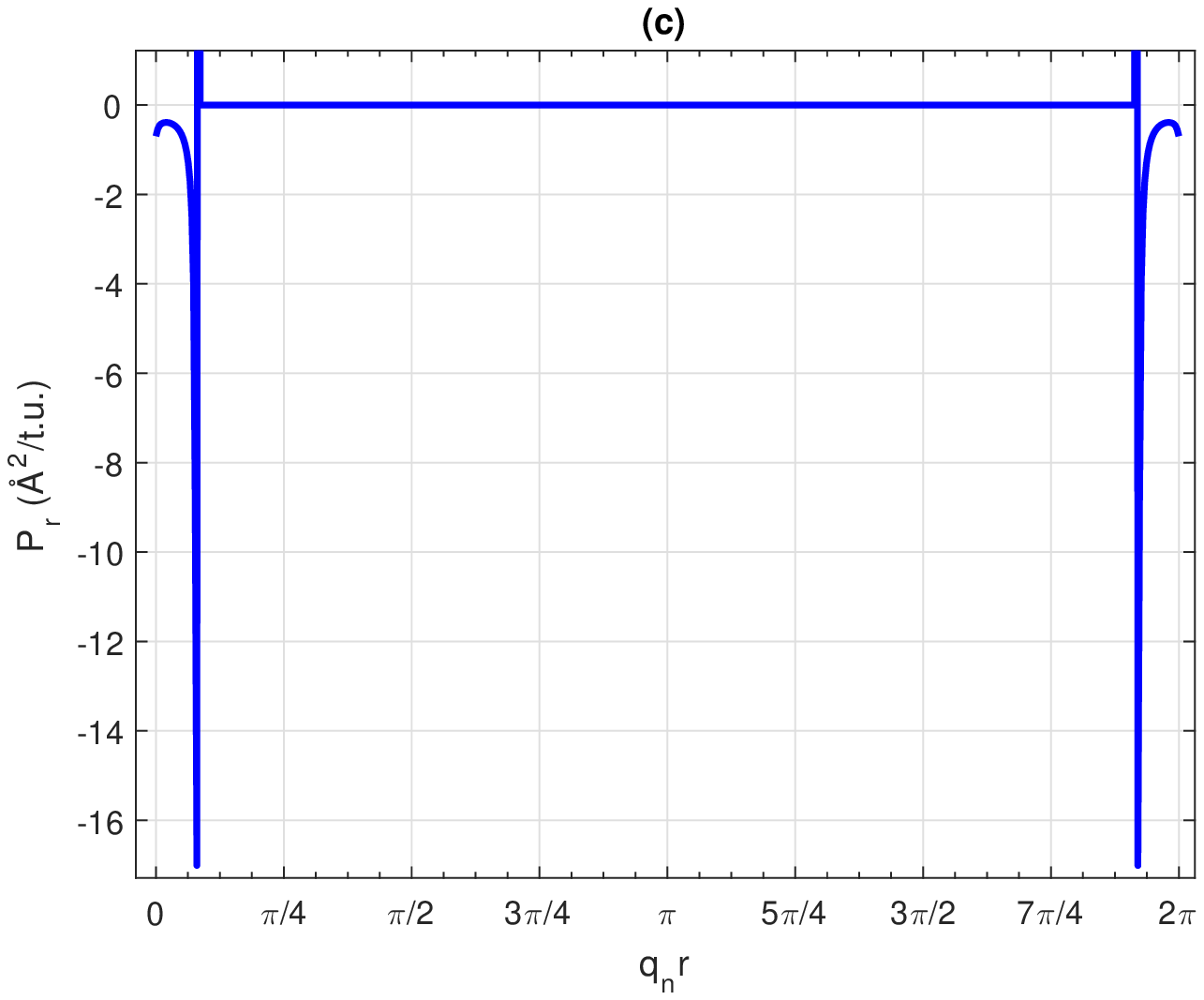}
\includegraphics[width=3in]{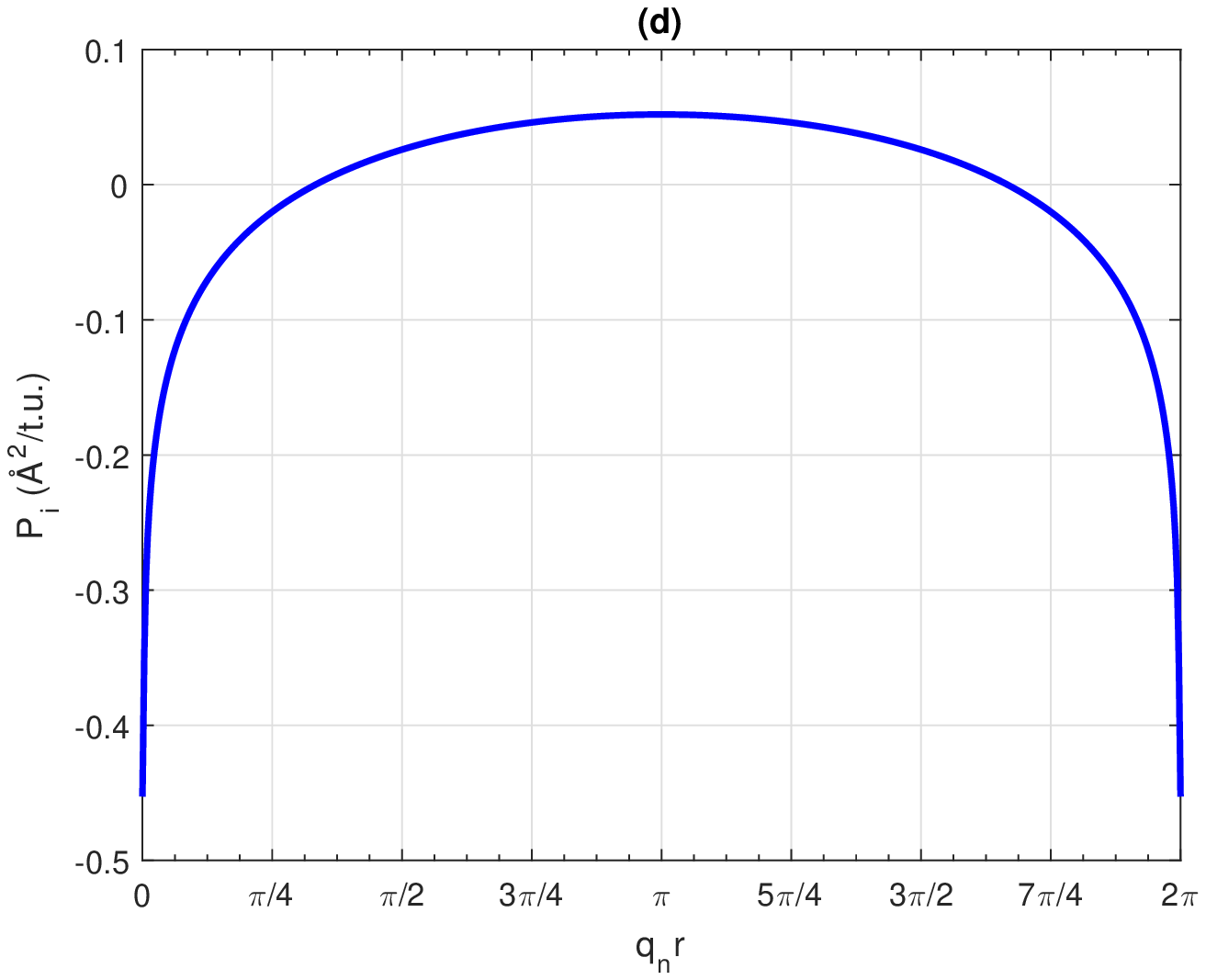}
\caption{(Color online) (a) and (b) the real and imaginary part of the dispersion relation (\eqref{eq17})
in the linear limit $F_1\rightarrow0$. (c) and (d) the real and imaginary part of the second discrete derivative of the
dispersion relation (\eqref{eq32}) (Dispersion coefficient) in terms of the discretized
values of the wave vector $q_nr=2\pi n/N$, $N=600$ for $s=3.00$, $\gamma_0=0.15$ $t.u^{-1}$.}\label{fig1}
\end{figure}

%%% -------------------------------------------------------------------
In this section, the DDO technique which is appropriate for
long-range interacting systems \cite{mil} is used to study the
dynamics of DNA breathing. Assuming as usual small amplitude
oscillation of the nucleotide around the bottom of the Morse
potential, we obtain up to the third order of the Morse potential
the following equation of motion:
\begin{equation}\label{eq13}
\begin{split}
\ddot v_n = &\sum\limits_{l=1}{\frac{J_l}{m}}(v_{n+l}
-2v_n + v_{n - l})-\omega_g^2(v_n+\alpha v_n^2+\beta v_n^3)-\gamma ^{St}\dot v_n +\sum\limits_{l = 1}
\gamma_l\left(\dot v_{n +l} - 2\dot v_n + \dot v_{n - l}  \right).
\end{split}
\end{equation}
where $\omega_g^2=\frac{4a^2D}{m}$, $\alpha=-\frac{3a}{\sqrt2}$ and $\beta=\frac{7a^2}{3}$. Eq. (\ref{eq13}) describes the dynamics of the out-of-phase motion of the DNA in viscous medium in the presence of LRI forces. Introducing the distance of neighboring bases $r$ and assuming plane wave solutions of the form
\begin{equation}\label{eq14}
\begin{split}
v_n =F_1e^{i(qnr-\omega t)}+ c.c.
\end{split}
\end{equation}
and substituting them into the equations of motion, we obtain the nonlinear dispersion relation in rotating wave approximation for the normal mode frequencies $\omega_{n}$ and wave numbers $q_{n}$
\begin{equation}\label{eq15}
\omega_{n}^{2} = \omega_g^2(1+3\beta|F_1|^2)+4\sum\limits_{l = 1}\frac{J_l}{m}
\sin^2(q_n^0lr/2)-i\omega_{n} \gamma_n,
\end{equation}
where $\gamma_n$ is the damping coefficient given by:
\begin{equation}\label{eq16}
\gamma_n=\gamma^{st}+4\sum\limits_{l=1}\gamma_l\sin^2(q_n^0rl/2).
\end{equation}
After some algebras, this dispersion relation can be rewritten as the
sum of its real part $\omega_r$ and imaginary part $\omega_i$. That
is
\begin{equation}\label{eq17}
\begin{split}
\omega_{n}=\omega_r+i\omega_i, \qquad \omega_r=\omega_{0,n}\sqrt{1-\delta_n^2},\qquad \omega_i=-\frac{\gamma_n}{2},\\
\end{split}
\end{equation}
with $\delta_n=\frac{\gamma_n}{2\omega_{0,n}}$ and $\omega_{0,n}$ the optical frequency of vibrations of base pairs in the absence of damping forces given by
\begin{equation}\label{eq18}
\begin{split}
\omega_{0,n}^2=\omega_g^2(1+3\beta|F_1|^2)+4\sum\limits_{l=1}\frac{J_l}{m}\sin^2(q_n^0rl/2).
\end{split}
\end{equation}
We plot in \figref{fig1} the real and imaginary parts of the angular frequency of the wave (\eqref{eq17}), the real and imaginary parts of the dispersion coefficient in the linear limit $|F_1|\rightarrow 0$ for discretized values of the wave vector $q_n^0r$ for $s =3.00$ and $\gamma_0=0.15$. In the panel (a) we observe that the real part of the angular frequency is equal to zero for $q_nr\in]\frac{\pi}{12}, \frac{23\pi}{12}[$ and different from zero otherwise, namely $q_nr\in[0, \frac{\pi}{12}]$ and $q_nr\in[\frac{23\pi}{12}, 2\pi]$. Then the vibration can appears and propagates in the DNA molecule only if the carrier wave vector $q_nr$ is selected in a finite interval $\Big\{[0, \frac{\pi}{12}]$ $\cup$ $[\frac{23\pi}{12}, 2\pi]\Big\}$.

The contribution of the long range decays of the stacking and viscous interactions are not the same, since the origins of the two forces are physically different, but nevertheless for seek of simplicity the same exponent is assumed that is $s=s'$. Also, as in \cite{are2}, we set $\gamma_0=\gamma^{st} =\gamma^{hy}$.
At small wavelength, the soliton solution of \eqref{eq13} is found as an expansion in normal modes and may be found in the form \cite{boy1}.
\begin{equation}\label{eq21}
\begin{split}
v_n =\varepsilon[F_1(z_1, \tau)e^{i(q^0rn-\omega^0t)}+ c.c.],
\end{split}
\end{equation}
with
\begin{equation}\label{eq22}
\begin{split}
F_1(z_1, \tau)=\sum\limits_{n=1}^N B_ne^{i(\delta q_nz_1-\delta\omega_{n}\tau)},\qquad q_n=q^0+\varepsilon\delta q_n, \qquad \omega_{n}=\omega^0+\varepsilon^2\delta\omega_{n}.
\end{split}
\end{equation}
 The function $F_1$ is a slowly varying function in space  $z_1=\varepsilon rn$ and time $\tau=\varepsilon^2t$. The parameter $q^0$ is the wavenumber of the wave packet and the associate frequency  $\omega^0\equiv\omega(q^0, F_1=0)$ in the limit $F_1\rightarrow 0$

The time derivative of the wave amplitude \eqref{eq22} reads:
\begin{equation}\label{eq23}
\begin{split}
\frac{\partial F_1(z_1, \tau)}{\partial\tau}=[-i\delta\omega_{n}] \sum\limits_{n=1}^N B_ne^{i(\delta q_nz_1-\delta\omega_{n}\tau)}=[-i\delta\omega_{n}]F_1.
\end{split}
\end{equation}
From Eqs. (\ref{eq15}), (\ref{eq22}) and (\ref{eq21}) it is seen that the term $\varepsilon^2\delta\omega_{n}$ is an evolution
function of two variables: the wavenumber $q^0$ and the slowly varying wave amplitude $|\varepsilon F_1|^2$. The Taylor expansion of this term
around the value $q^0$ and $|\varepsilon F_1=0|$ and neglecting higher order terms $(>2)$, give us
\begin{equation}\label{eq24}
\begin{split}
\varepsilon^2\delta\omega_{n}(\partial q_n,|\varepsilon F_1|^2)=\delta\omega^0(q^0)+
(q_n-q^0)\frac{\partial \omega^0(q^0)}{\partial q^0}
+\frac{1}{2}(q_n-q^0)^2\frac{\partial^2 \omega^0(q^0)}{(\partial {q^0})^2}
+|\varepsilon F_1|^2\frac{\partial \omega_{n}(q^0)}{\partial(|F_1|^2)}\Big|_{F_1=0}.
\end{split}
\end{equation}
The first term of the right hand site of \eqref{eq24} is assuming to be very close to zero. From \eqref{eq22} we have $\varepsilon\delta q_n=q_n-q^0$.
The above considerations in \eqref{eq24} lead to:
\begin{equation}\label{eq25}
\begin{split}
\delta\omega_{n}(\partial q_n,|F_1|^2)=\frac{(\varepsilon\delta q_n)}{\varepsilon^2}\frac{\partial \omega^0(q^0)}{\partial q^0}
+\frac{1}{2}\frac{(\varepsilon\delta q_n)^2}{\varepsilon^2}\frac{\partial^2 \omega^0(q^0)}{\partial {q^0}^2}
+|F_1|^2\frac{\partial \omega_{n}(q^0)}{\partial(|F_1|^2)}\Big|_{F_1=0}.
\end{split}
\end{equation}
The discrete difference operator is used instead of the continuous derivatives which can cause divergences. Therefore we have:
\begin{equation}\label{eq26}
\begin{split}
\frac{\partial \omega^0(q^0)}{\partial q^0}=\frac{\omega^0(q^0+h)-\omega^0(q^0)}{h},
\qquad \frac{\partial^2\omega^0(q^0)}{\partial {q^0}^2}=\frac{\omega^0(q^0+h)-2\omega^0(q^0)+\omega^0(q^0-h)}{h^2},
\end{split}
\end{equation}
and finally we get,
\begin{equation}\label{eq27}
\begin{split}
\delta\omega_{n}(\partial
q_n,|F_1|^2)=\sum\limits_{\nu=1}^2\frac{(\varepsilon\delta
q_n)^\nu}{\varepsilon^2\nu!}
\frac{\Delta_h^{(\nu)}[\omega^0(q^0)]}{h^\nu}+|F_1|^2\frac{\partial
\omega_{n}(q^0)}{\partial(|F_1|^2)}\Big|_{F_1=0},
\end{split}
\end{equation}
where $\Delta^{(\nu)}_h$ is the difference operator of order $\nu$ with step size $h=2\pi/N$ in the limit $F_1=0$ and given below,
\begin{equation}\label{eq28}
\begin{split}
\Delta^{(1)}_h[\omega^0]=\omega^0(q^0+h)-\omega^0(q^0), \qquad
\Delta^{(2)}_h[\omega^0]=\omega^0(q^0+h)-2\omega^0(q^0)+\omega^0(q^0-h).
\end{split}
\end{equation}
From \eqref{eq22} the term $(\delta q_n)^\nu$ can be expressed as follows:
\begin{equation}\label{eq29}
\begin{split}
(i\delta q_n)^\nu F_1=\frac{\partial^\nu F_1}{\partial z_1^\nu}.
\end{split}
\end{equation}
Using Eqs. [\ref{eq29}-\ref{eq26}] into \eqref{eq23} give the nonlinear equation of evolution of the envelope
function written as
\begin{equation}\label{eq30}
\begin{split}
i\Big[\frac{\partial F_1}{\partial \tau}+\frac{v_g}{\varepsilon}\frac{\partial F_1}{\partial z_1}\Big]+P\frac{\partial^2F_1}{\partial z^2}+Q|F_1|^2F_1=0
\end{split}
\end{equation}
where the parameters $v_{g}$, $P$ and $Q$ are the group velocity, the dispersion and the nonlinearity coefficients given by
\begin{equation}\label{eq31}
\begin{split}
v_{g}=\frac{\Delta^{(1)}_h[\omega^0]}{h},
\qquad
P=\frac{\Delta^{(2)}_h[\omega^0]}{2h^2},
\qquad
Q=-\frac{\partial \omega_{n}(q^0)}{\partial(|F_1|^2)}\Big|_{F_1=0}.
\end{split}
\end{equation}
The above parameters can be rewritten as:
\begin{equation}\label{eq32}
\begin{split}
&v_{g}=v_{gr}+iv_{gi},\qquad v_{gr}=\frac{\Delta^{(1)}_h[\omega_r^0]}{h},\qquad v_{gi}=\frac{\Delta^{(1)}_h[\omega_i^0]}{h}\\
&P=P_{r}+iP_{i}, \qquad P_{r}=\frac{\Delta^{(2)}_h[\omega_r^0]}{2h^2},\qquad P_{i}=\frac{\Delta^{(2)}_h[\omega_i^0]}{2h^2},\\
&Q=Q_{r}+iQ_{i}, \qquad Q_{r}=-\frac{3\beta\omega_g^2}{2\omega_r^0},\qquad Q_{i}=0.
\end{split}
\end{equation}
Setting $\xi_1=z_1-\varepsilon v_{g}t$ in the co-moving
reference frame with a rescaled time $t$ such as $t\rightarrow\varepsilon^2t$,
\eqref{eq27} becomes
\begin{equation}\label{eq33}
\begin{split}
i\frac{\partial F_1}{\partial t}+(P_{r}+iP_{i})\frac{\partial^2F_1}{\partial \xi_1^2}+Q|F_1|^2F_1=0.
\end{split}
\end{equation}
It should be noted that \eqref{eq33} is the well-known CGL equation for the evolution of the envelope where the dispersion coefficient is complex and the nonlinearity coefficient is real. Similar equation was found in Ref. \cite{zdr2, zdr22} where the authors studied the dynamics of a damped DNA in the absence of hydrodynamical damping and LRI forces using the semi-discrete approximation. In their study they found the dispersion coefficient real and the nonlinearity coefficient complex contrary of the one obtained in this work. The nonlinearity coefficient $Q$ and the dispersion coefficient $P$ not only depend on the wave vector $q_n^0r$, and the Stokes viscous forces as previously mentioned by
these authors, but also depend on the hydrodynamic damping and the LRI forces.

Several methods related to soliton solutions for the specific forms of CGL equation have been developed \cite{Stenflo, hoc, akh}. A key problem in this paper is to give an analytical soliton solution of the CGL equation (\eqref{eq33}), and use it to study the effect of
viscosity and LRI on the DNA opening state configuration. The character of this solution is determined by the sign of $Q$ and $P_{r}$
while the stability of the plane wave solution through the Benjamin-Feir instability depends on the sign of the product $P_{r}Q$. For $P_{r} Q<0$,
the plane wave solution is stable and for $P_{r} Q>0$ it is unstable. Particularly, since the nonlinear coefficient $Q$ is always negative the sign of the constant $P_{r}Q$ depends on real part of the dispersion coefficient $P_{r}$ which can take positive or negative values depending on the range of variations of the wave vector. Here, only localized solutions in space for any wave carrier whose wavenumber is in the positive range of $P_rQ$ are considered.

\begin{figure}[H]
\centering
\includegraphics[width=3in]{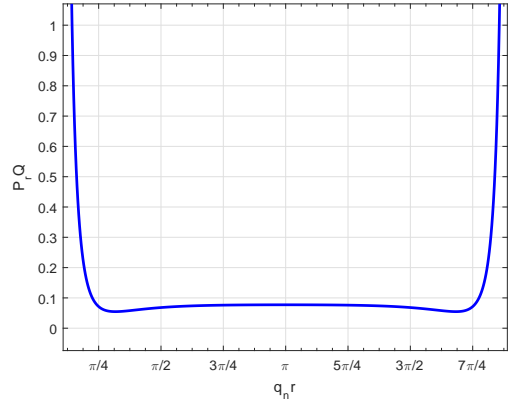}
\caption{(Color online) The product $PrQ$ in terms of the
the discretized values of the wave vector $q_nr=2\pi n/N$, $N=600$,
for $s=3.00$, $\gamma_0=0.15$ $t.u^{-1}$}\label{fig2}
\end{figure}

In \figref{fig2}, the product $P_rQ$  is represented as a function of the wave vector for $s=3.00$ and $\gamma=0.15$. We observe in the plots that $P_{r}Q>0$ is always positive. From Eqs. (\ref{eq16}), (\ref{eq17}) and (\ref{eq32}), we observe that the imaginary parts of the solitonic parameters
strongly depend on the damping forces of the system, therefore their absolute values decrease with the decreasing of the damping constant and vanish when the damping is switched off.

As in \cite{Stenflo, hoc, akh}, an analytical solution of \eqref{eq34}, is found in the form
\begin{equation}\label{eq34}
\begin{split}
&F_1=A\Big[\sech(\eta\xi_1)\Big]^{1+i\sigma}e^{-i\phi t},\\
\end{split}
\end{equation}
where $A$, $\phi$, $\eta^{-1}$, and $\sigma$ are parameters to be determined and represent respectively the complex
amplitude, the complex \textquotedblleft angular frequency \textquotedblright, the width  and the chirp of the soliton. By introducing Eq. (\ref{eq34}) into Eq (\ref{eq33}) and after canceling the terms in $[\sech(\eta\xi_1)\Big]^{1+i\sigma}$, the real and imaginary part of the phase of the soliton is written
\begin{equation}\label{eq35}
\begin{split}
&\phi_r=-\eta^2\Big[(1-\sigma^2)P_{r}-2\sigma P_{i}\Big],
\qquad \phi_i=-\eta^2\Big[(1-\sigma^2)P_{i}+2\sigma P_{r}\Big].
\end{split}
\end{equation}
From the annihilation of the terms in $[\sech(\eta\xi_1)\Big]^{3+i\sigma}$, the width and the chirp of the soliton is given by
\begin{equation}\label{eq36}
\begin{split}
&|A|=\eta\sqrt{\Big|\frac{(2-\sigma^2)P_{r}-3\sigma P_{i}}{Q_{r}}}\Big|,
\qquad \sigma=\frac{3P_{r}+\sqrt{\Delta}}{2P_{i}}\\
\end{split}
\end{equation}
where $\Delta=9P_{r}^2+8P_{i}^2$ and $\sigma$ a solution of the following quadratic equation
\begin{equation}\label{eq37}
\begin{split}
&P_{i}\sigma^2-3P_{r}\sigma-2P_{i}=0.
\end{split}
\end{equation}
This choice of $\sigma$ implies that the soliton is strongly chirped.

Now to determine the complex amplitude $A_{\gamma}$ of the soliton, let us consider the system in the non-viscous limit ($\gamma_0=0$). In that case $P_{i}$ vanishes and Eq. (\ref{eq33}) becomes the standard NLS equation
\begin{equation}\label{eq38}
i\frac{\partial G_1}{\partial t}+P'\frac{\partial
^2G_1}{\partial\xi_1^2}+Q'|G_1|^2G_1=0,
\end{equation}
where the associated group velocity, dispersion coefficient and nonlinearity coefficient are
\begin{equation}\label{eq39}
\begin{split}
v_g'\equiv v_{gr}\Big|_{\gamma=0}=\frac{\Delta^{(1)}_h[\omega_0^0]}{h},
\qquad P'\equiv P_{r}\Big|_{\gamma=0}=\frac{\Delta^{(2)}_h[\omega_0^0]}{2h^2},
\qquad Q'\equiv Q_{r}\Big|_{\gamma=0}=-\frac{3\beta\omega_g^2}{2\omega_0}.
\end{split}
\end{equation}
The solution of Eq. (\ref{eq38}) is the well known modulated solitonic wave called breather \cite{pey, sco} and given by
\begin{equation}\label{eq40}
\begin{array}{l}
G_1=A'\sech\Big[L(\xi_1-u_e t)\Big]e^{i\frac{u_e}{2P'}(\xi_1-u_ct)},
\end{array}
\end{equation}
where $u_e$ and $u_c$ are real parameters representing respectively the velocities of the envelope and the carrier wave of the soliton. The amplitude of the envelope  $A'$ and its  inverse width $L'$ are given by the relations
\begin{equation}\label{eq41}
\begin{split}
L'=\frac{\sqrt{u_e^2-2u_eu_c}}{2P'},\qquad A'=\sqrt{\frac{u_e^2-2u_eu_c}{2P'Q'}}.
\end{split}
\end{equation}

Let us assume that at the initial time ($t=0$), for $\gamma_0=0$, the soliton solution Eqs. (\ref{eq34}) and (\ref{eq41}) should be equivalent,
\begin{equation}\label{eq42}
\begin{array}{l}
A\Big|_{\gamma=0}=A'e^{i\frac{u_e}{2P'}\xi_{1,0}},
\end{array}
\end{equation}
where $\xi_{1,0}$ is the initial position of the soliton. Taking into account this new expression of the amplitude $A$, the one soliton solution of \eqref{eq33} is
\begin{equation}\label{eq43}
\begin{split}
&F_1=\eta\sqrt{\Big|\frac{(2-\sigma^2)P_{r}-3\sigma P_{i}}{Q_{r}}}\Big|e^{\phi_it}\Big[\sech(\eta\xi_1)\Big]^{1+i\sigma}e^{i(\frac{u_e}{2P}\xi_{1,0}-\phi_r t)}.
\end{split}
\end{equation}
Now considering the fact that the angular frequency of the soliton is complex (see Eq. (\ref{eq17})), one can note that the term $e^{\omega_it}$ enters inside the amplitude of the soliton. Thus, the complete expression of the envelope soliton written in the original temporal ($t$) and spacial ($z$) variables is finally given by
\begin{equation}\label{eq44}
\begin{split}
F'_1=&F_1e^{\omega_it}=\eta\sqrt{\Big|\frac{(2-\sigma^2)P_{r}-3\sigma P_{i}}{Q_{r}}}\Big|e^{-\Gamma t}\Big[\sech\big[\eta(z_1-\varepsilon v_{gr}t)\big]\Big]^{1+i\sigma}e^{i\Theta}\\
\end{split}
\end{equation}
with
\begin{equation}\label{eq45}
\begin{split}
\Theta=&\frac{u_e}{2P}\xi_{1,0}+\eta^2\Big[(1-\sigma^2)P_{r}-2\sigma P_{i}\Big]t, \qquad
\Gamma=\frac{\gamma_n}{2}+\eta^2\Big[(1-\sigma^2)P_{i}+2\sigma P_{r}\Big], \qquad
\eta=\frac{\sqrt{u_e^2-2u_eu_c}}{\Big|(2-\sigma^2)P_{r}-3\sigma P_{i}\Big|}
\end{split}
\end{equation}
where $\Gamma$ is the effective damping constant of the medium.

\begin{figure}[H]
\centering
\includegraphics[width=3in]{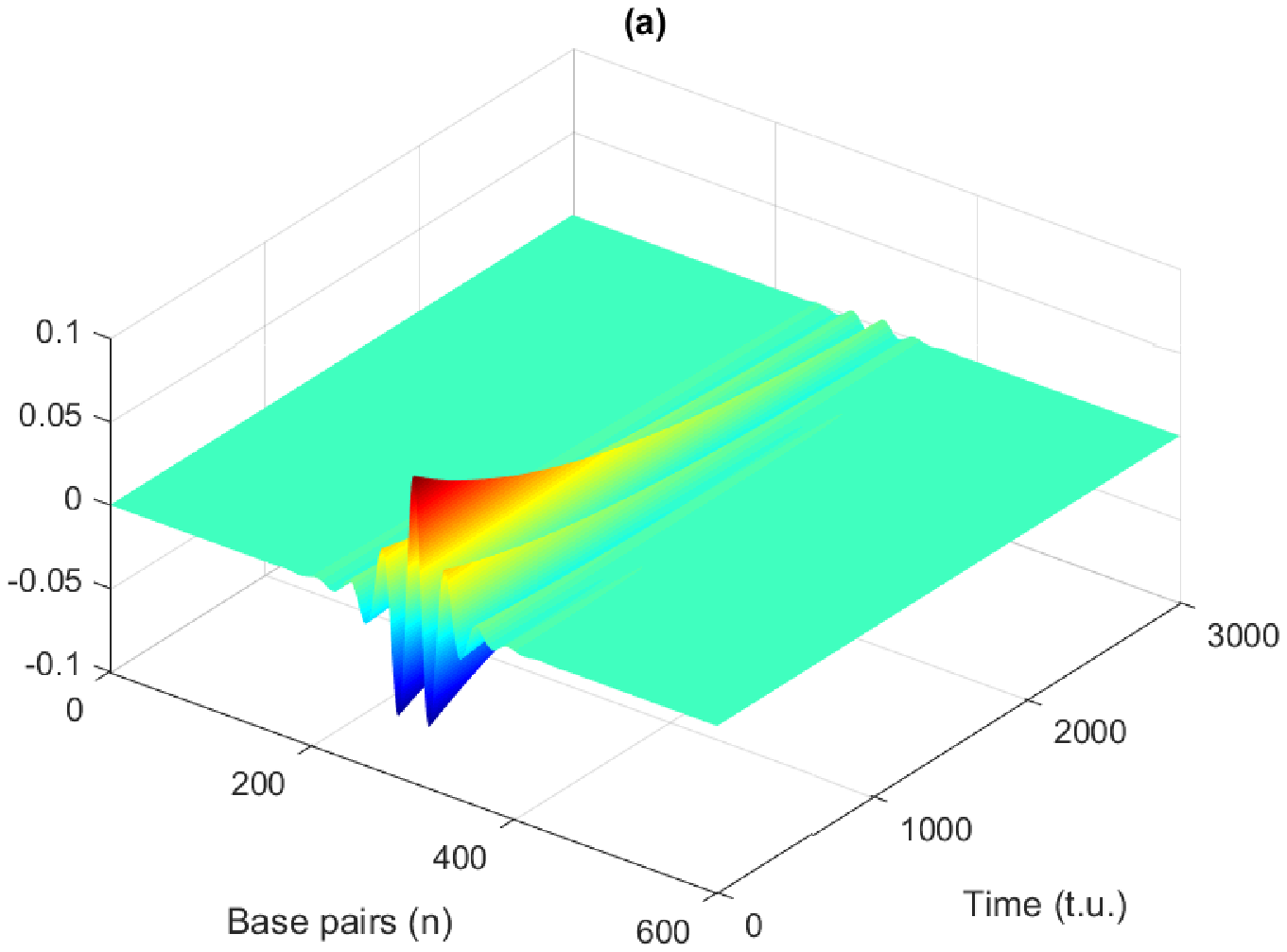}
\includegraphics[width=3in]{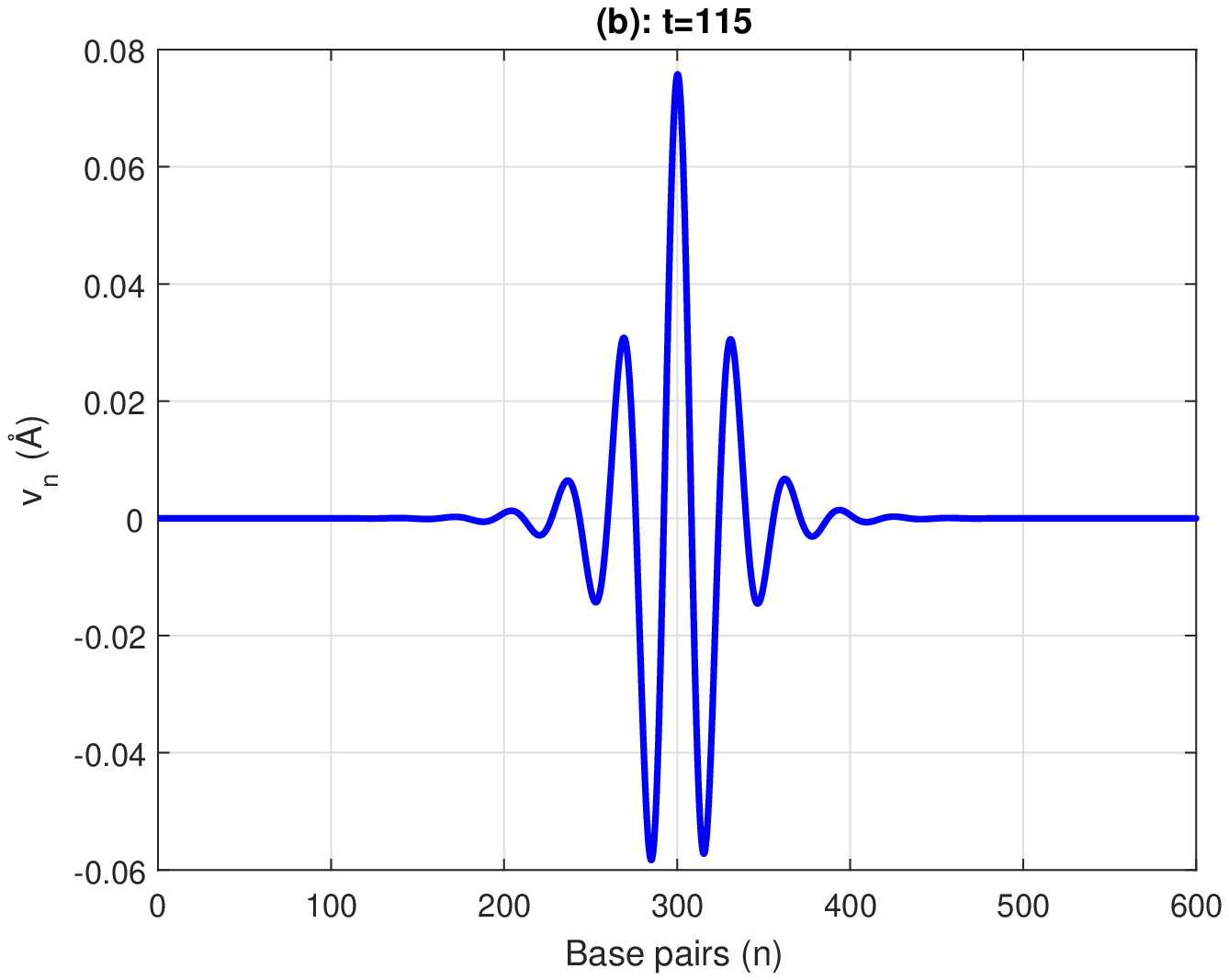}
\caption{(Color online) (a) Analytical stretching of the nucleotide pair as a function of the time and the number of base pairs. (b) stretching of the nucleotide pair as a function of the number of base pairs at $t=115$ for $s=3.00$, $\varepsilon=0.9$, $u_e=1$, $\gamma_0=0.15$ $t.u^{-1}$, $u_c=0.45u_e$ and $q_n^0r=\frac{\pi}{16}$.}\label{fig3}
\end{figure}

To obtain the solution of the EOM of the out-of-phase motion $v_n$ given by Eq. (\ref{eq12}), some results of the previous section are used. Using Eqs. (\ref{eq17}), (\ref{eq31}) and (\ref{eq43}) the soliton solution describing the out-of-phase motion of the DNA in the viscous medium takes the form
\begin{equation}\label{eq46}
\begin{split}
v_n(t)=&2\varepsilon |A| e^{-\Gamma t}\sech\big[\varepsilon\eta(nr-v_{gr}t)\big]\cos(q_{\gamma}nr-\varpi t)
\end{split}
\end{equation}
where
\begin{equation}\label{eq47}
\begin{split}
&\varpi=\omega_r^0-\eta^2\Big[(1-\sigma^2)P_{r}-2\sigma P_{i}\Big],
\qquad
q_{\gamma}=q^0+\frac{u_e}{2P}\Big(\frac{\xi_{1,0}}{nr}\Big)+(\sigma/nr)\log\Big|\sech\big[\varepsilon\eta(nr-v_{gr}t)\big]\Big|.\\
\end{split}
\end{equation}
The solution \eqref{eq46} represents the breather solution in the DNA molecule suggested by the Infrared and Raman experiments \cite{pro1}. This solution is represented in \figref{fig3}. It can be seen that due to the damping effect, the amplitude of the soliton is a decreasing function of time and hence it will propagates only for a limited distance and vanishes.

Prohofsky \emph{et al.} have shown in their studies \cite{pro1} that, the breather can be strongly located, or distributed on a wide zone of the DNA molecule and could be at the origin of the localization of the energy in the molecule which lead to the local denaturation.

It is noteworthy to mention that if one uses the semi-discrete approach \cite{zdr2, zdr22}, the results fails in the derivation of the breather soliton profile \eqref{eq46}. In fact, we can obtain a similar CGL equation \eqref{eq33}, but with a different dispersion coefficient containing continuous derivatives instead of difference operators as given in equation \eqref{eq31}. Using the semi-discrete approach, we obtain:

\begin{equation}\label{eq48}
\begin{split}
&P_{r}=\frac{1}{2\omega_r} \Big[\sum\limits_{l=1}\Big(\frac{J_l}{m}+\omega_i\gamma_l\Big)(rl)^2\cos(qlr)-|v_{g
\gamma}|^2\Big],\;\ P_{i}=-\frac{1}{2}\sum\limits_{l=1}\gamma_l(rl)^2\cos(qrl),\\
&\omega_r=\omega\sqrt{1-\Big(\frac{\gamma}{2\omega}\Big)^2},\; \omega^2=\omega_g^2+4\sum\limits_{l=1}\frac{J_l}{m}\sin^2(qrl/2),\;\ \omega_i=-\frac{\gamma}{2},\;\ \gamma=\gamma^{st}+4\sum\limits_{l=1}\gamma_l\sin^2(qrl/2),\\
&v_{gr}=\frac{r}{\omega_r}\sum\limits_{l=1}\Big(\frac{J_l}{m}+\omega_i\gamma_l\Big)l\sin(qrl),\;\ v_{gi}=-r\sum\limits_{l=1}\gamma_ll\sin(qrl),\;\ \varsigma=\omega_g^2+\frac{4}{m}\sum\limits_{l =1}J_l\sin^2(qrl),\\
&Q_{r}=-\frac{\omega_g^2\alpha}{\omega_r}(-2\alpha+\frac{3}{2}\beta/\alpha+BC),\;\  Q_{i}=-\frac{\omega_g^2\alpha}{\omega_r}CD_1\;\ B=4\omega_r^2\varsigma, \;\ D_1=2\gamma\omega_r,\;\ C=\frac{\omega_g^2\alpha}{B^2+D_1^2}.
\end{split}
\end{equation}
In \eqref{eq48}, the appearance of continuous derivatives can cause a divergence of both the group velocity $v_{g}$ and the dispersion coefficient $P$. The coefficients oscillate for different values of chain length and does not converge to a definite value.
\section{\label{solution}Numerical investigations.}
\begin{figure}[H]
\centering
\includegraphics[width=3in]{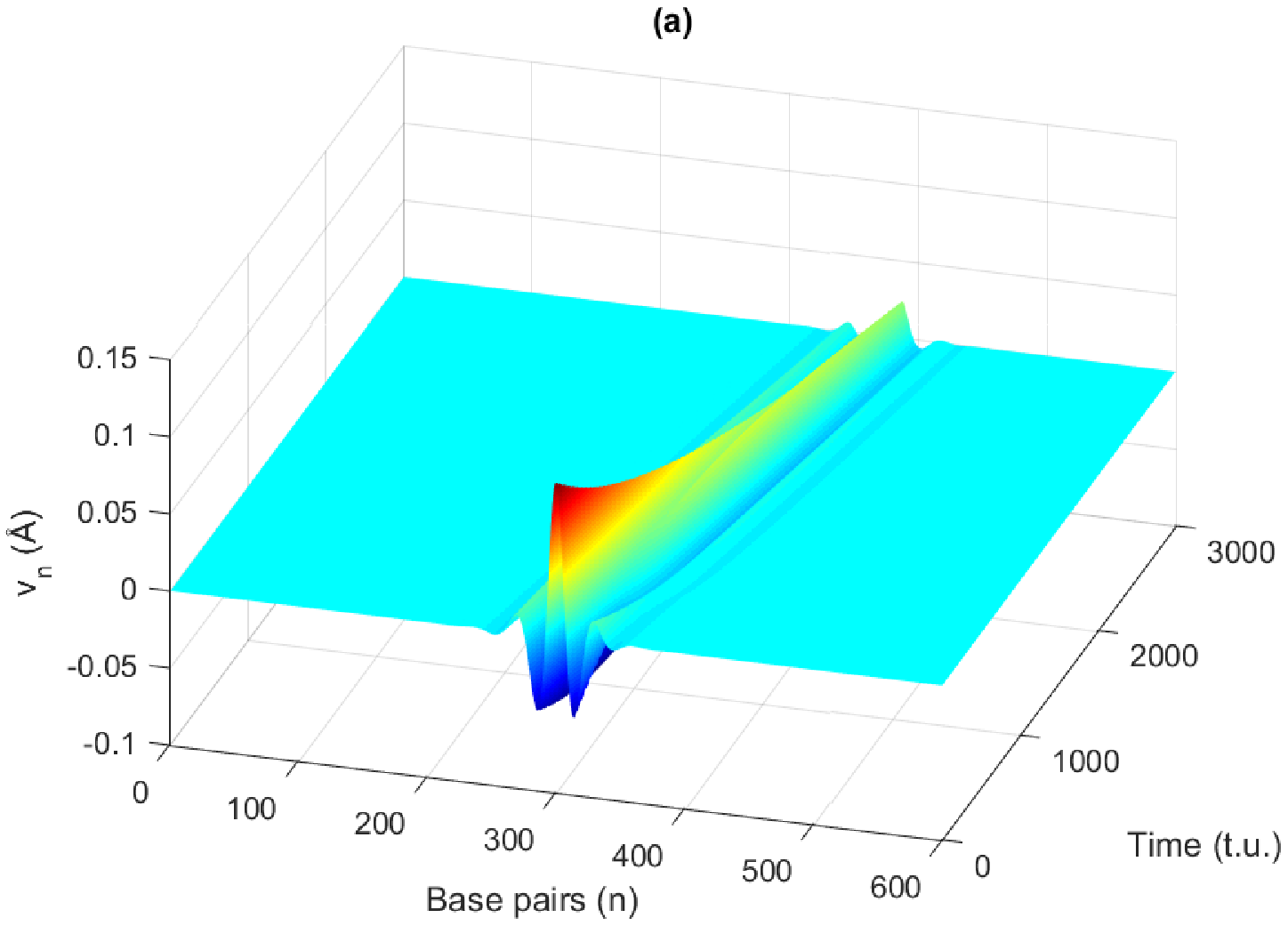}
\includegraphics[width=3in]{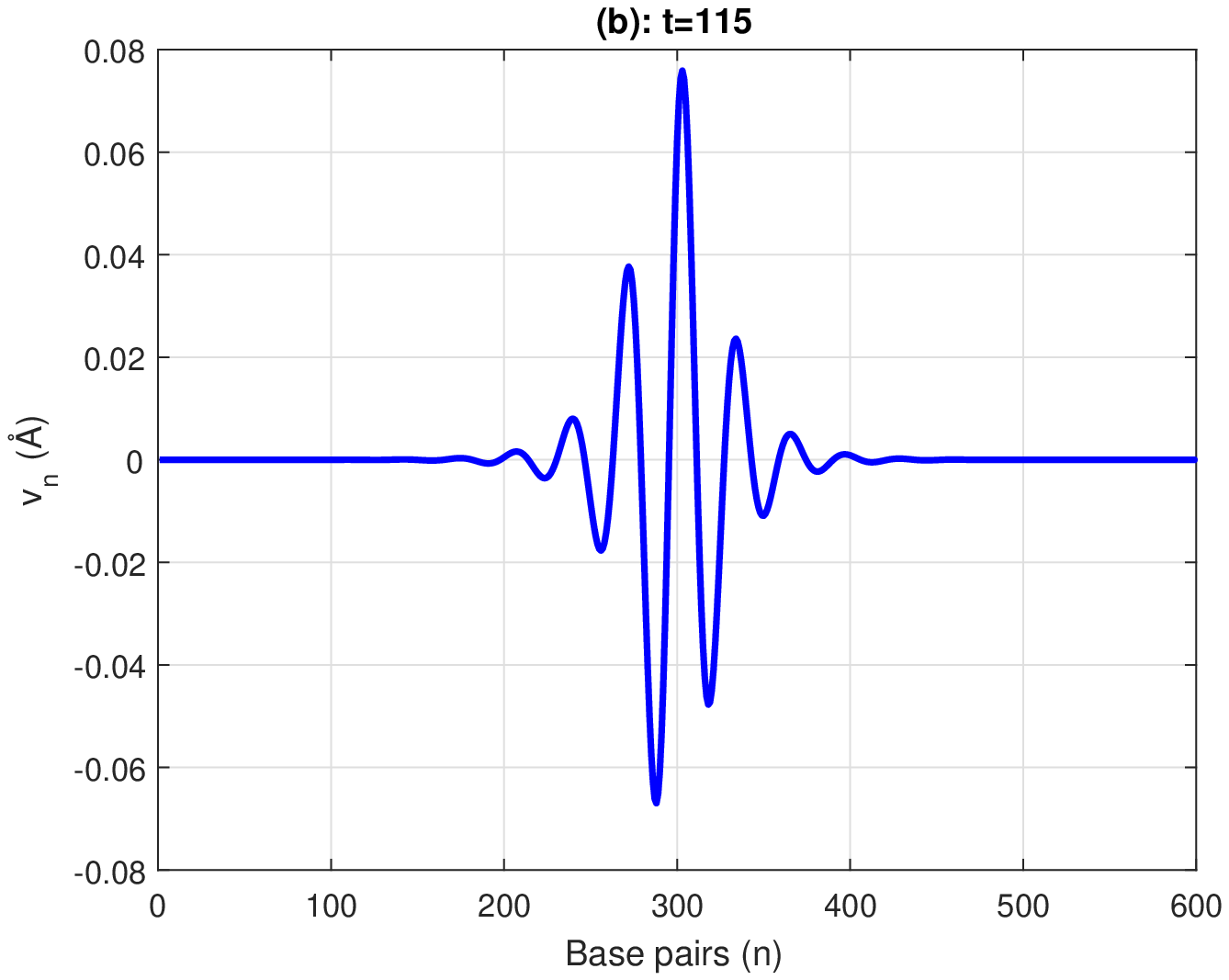}
\caption{(Color online) (a) Numerical stretching of the nucleotide pair as a function of the time and the number of base pairs. (b) stretching of the nucleotide pair as a function of the number of base pairs at $t=115$ for $s=3.00$, $\varepsilon=0.9$, $u_e=1$, $\gamma_0=0.15$ $t.u^{-1}$, $u_c=0.45u_e$ and $q_n^0r=\frac{\pi}{16}$.}\label{fig4}
\end{figure}

\begin{figure}[H]
\centering
\includegraphics[width=3in]{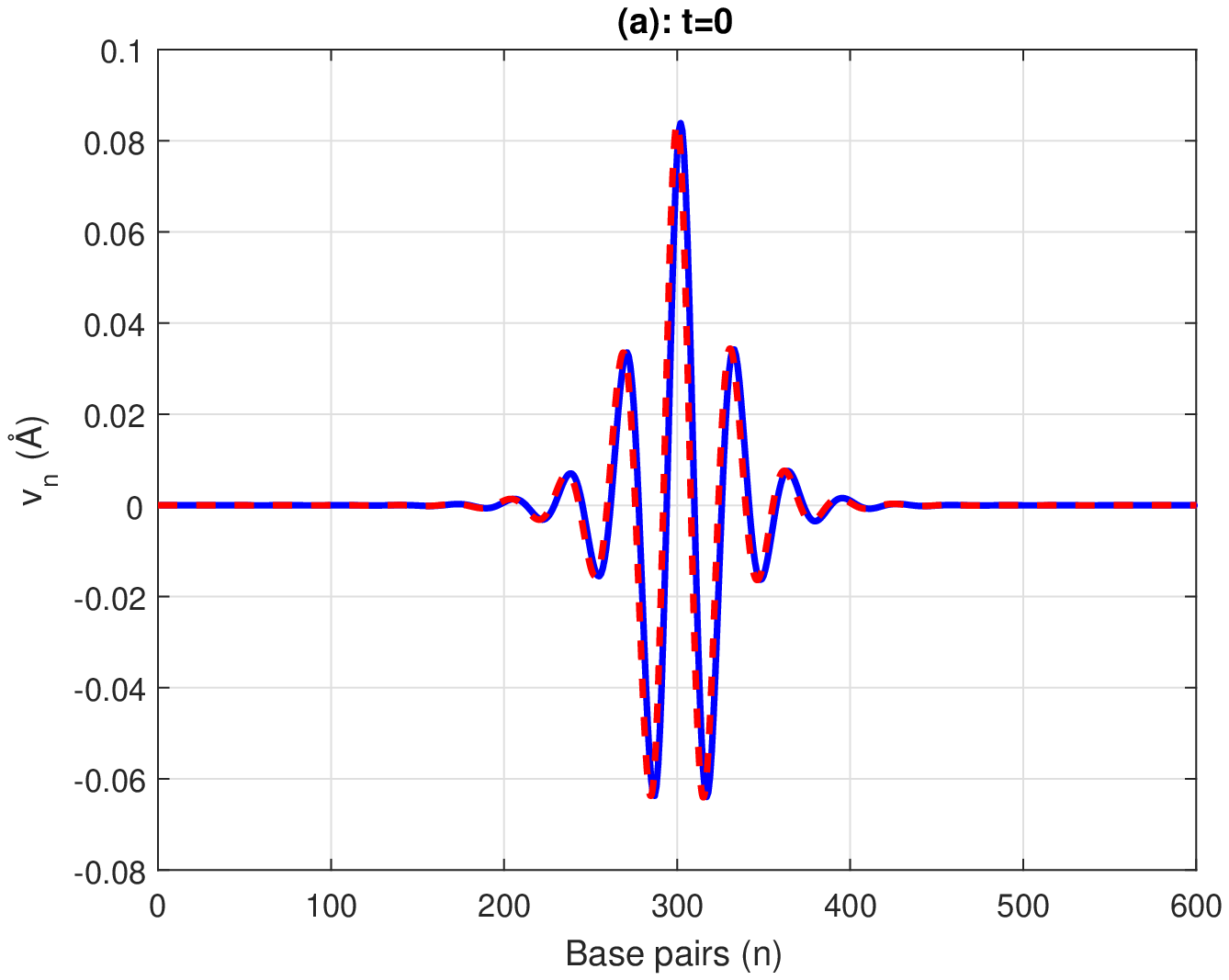}
\includegraphics[width=3in]{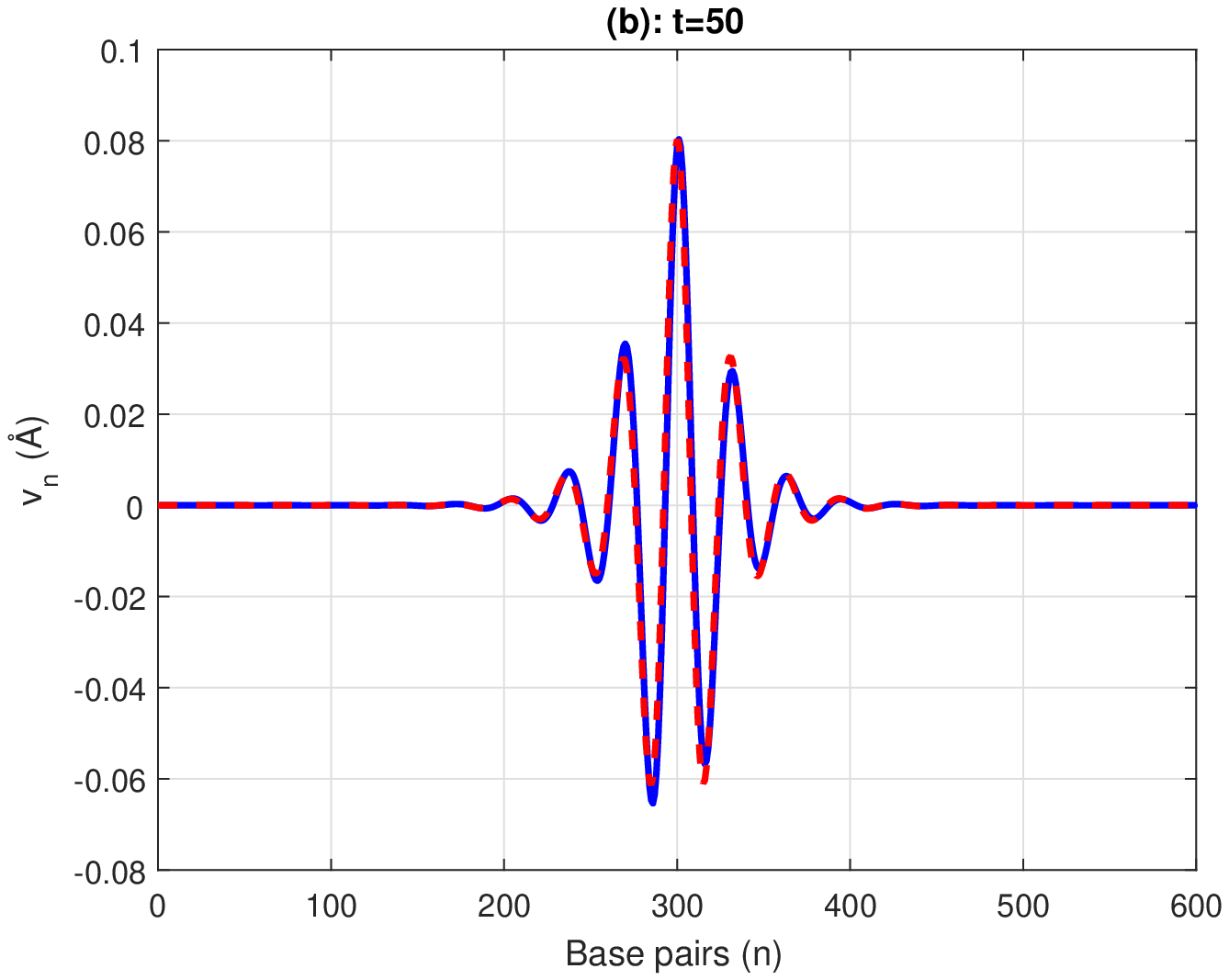}
\includegraphics[width=3in]{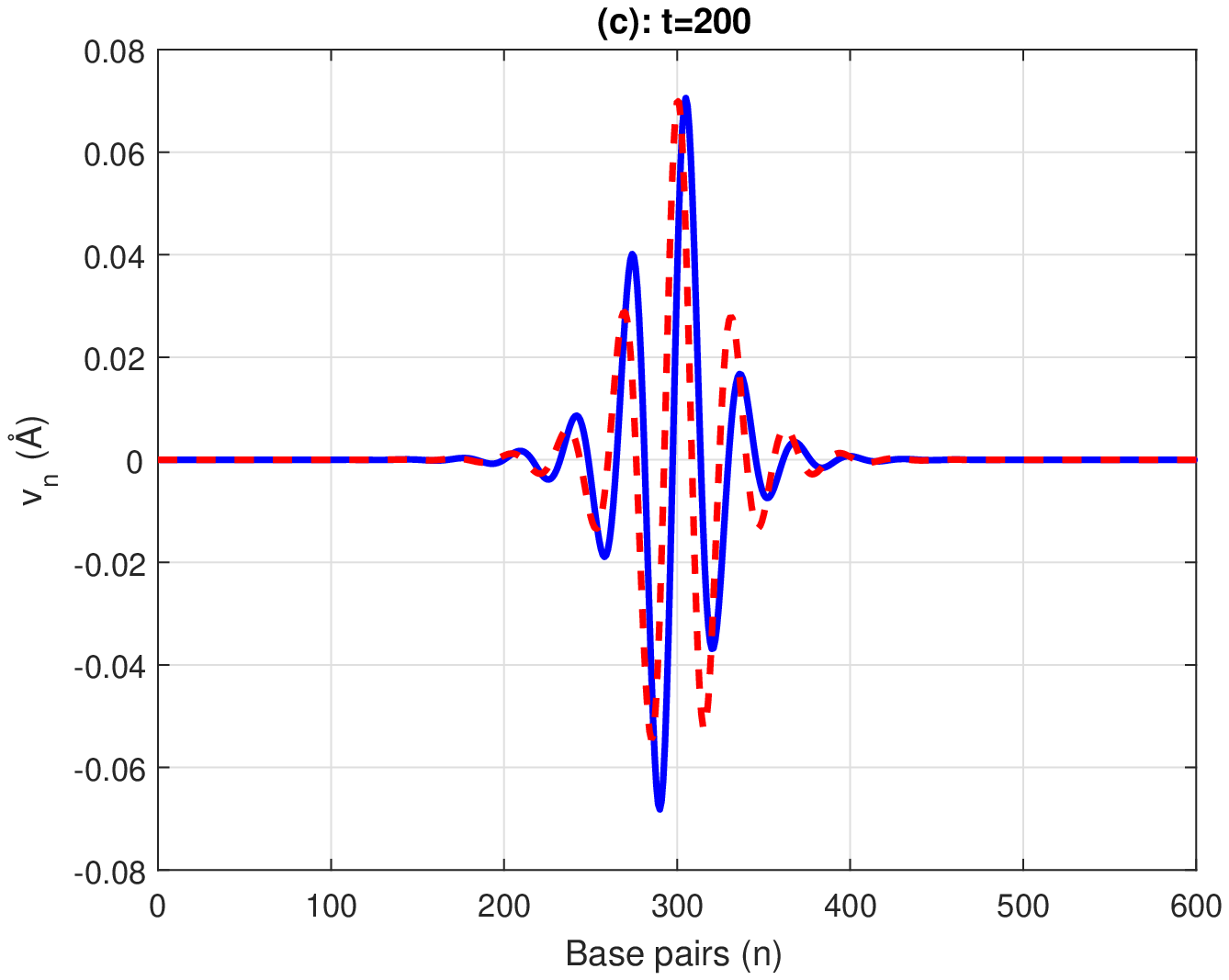}
\includegraphics[width=3in]{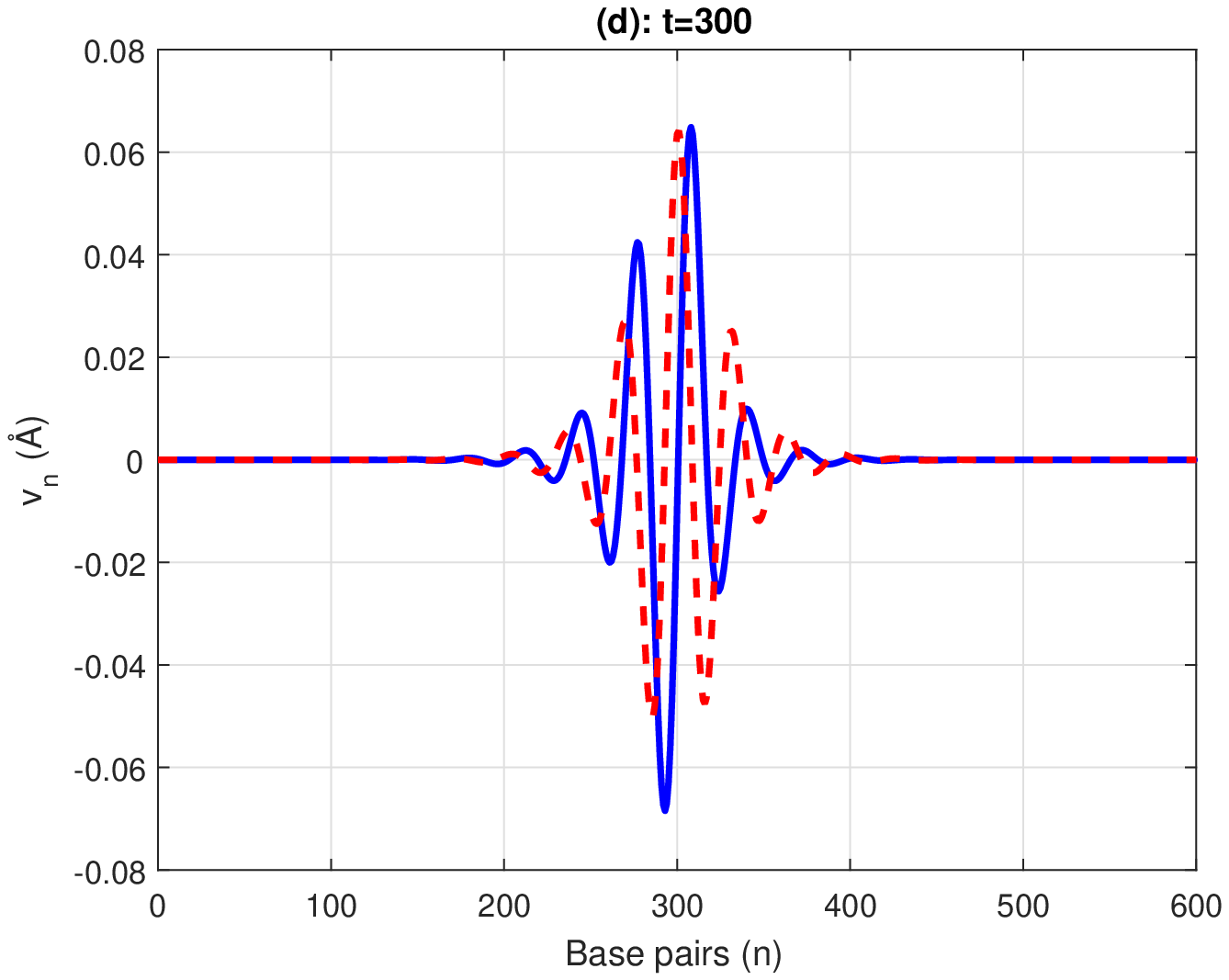}
\caption{(Color online) Comparison between analytical and  numerical solution of the EOM \eqref{eq12} at different time positions for $s=3.00$, $\varepsilon=0.9$, $u_e=1$, $\gamma_0=0.15$ $t.u^{-1}$, $u_c=0.45u_e$ and $q_n^0r=\frac{\pi}{16}$. (solid blue line) the numerical solution. (dash red line) the analytical solution.}\label{fig5}
\end{figure}

\begin{figure}[H]
\centering
\includegraphics[width=3in]{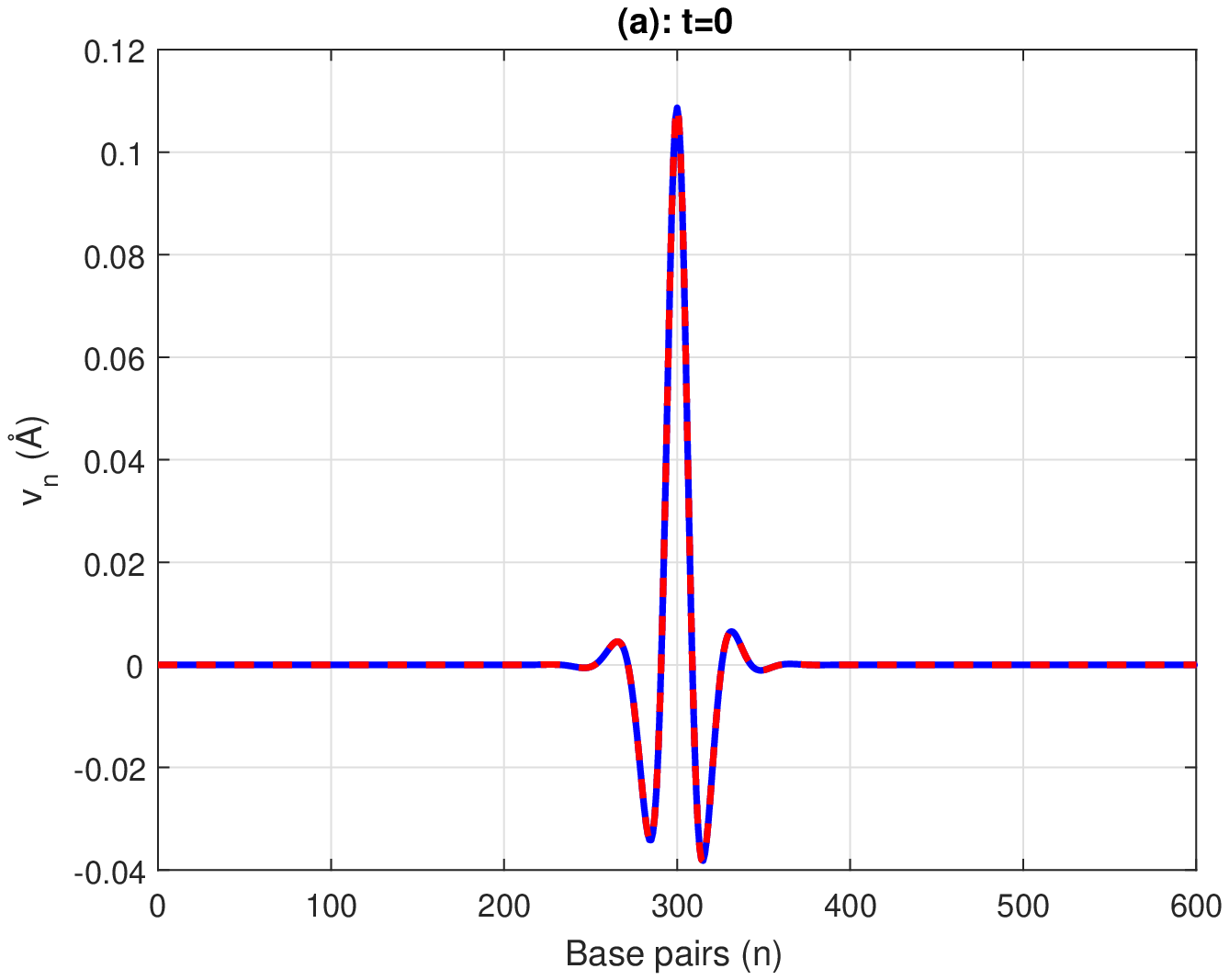}
\includegraphics[width=3in]{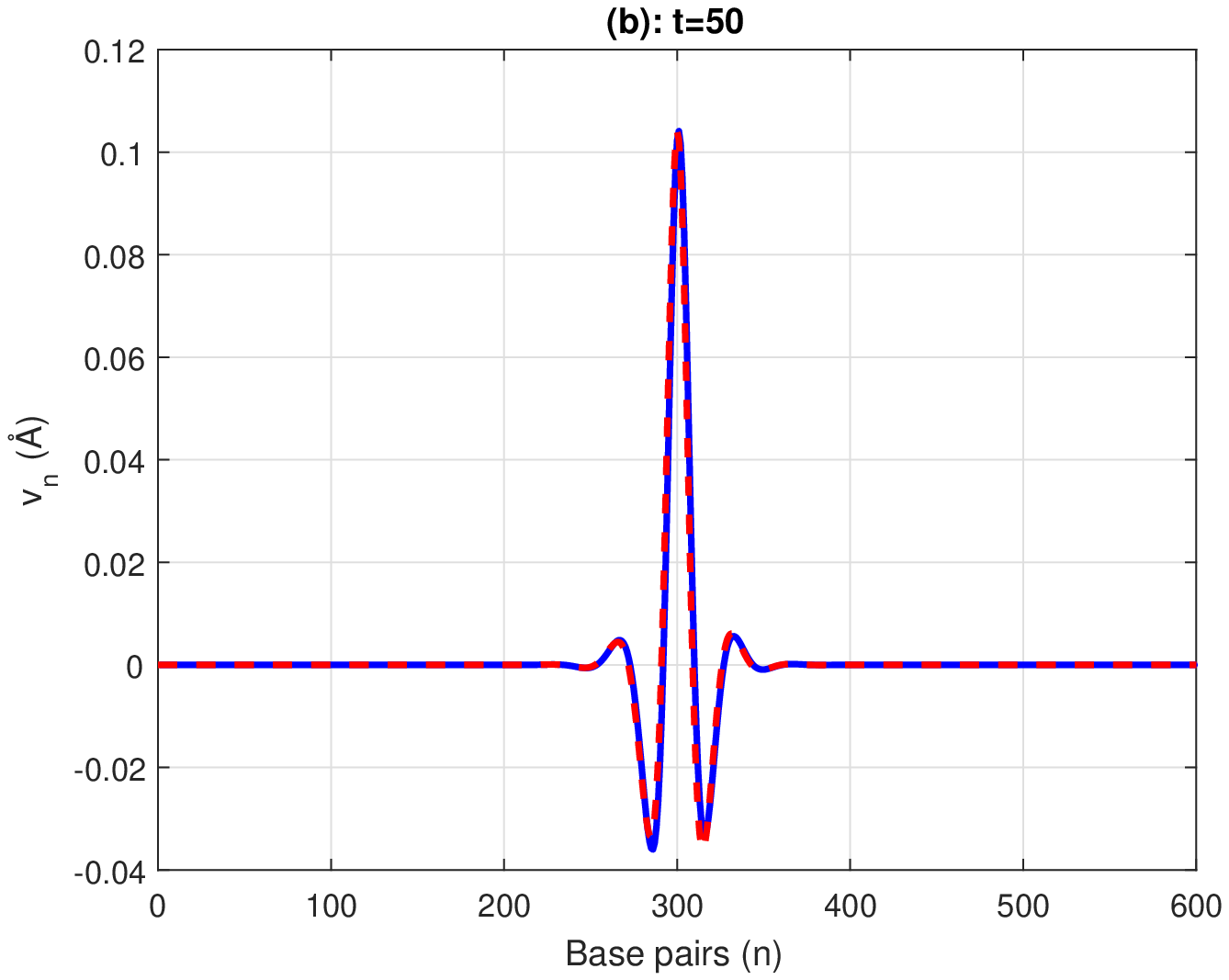}
\includegraphics[width=3in]{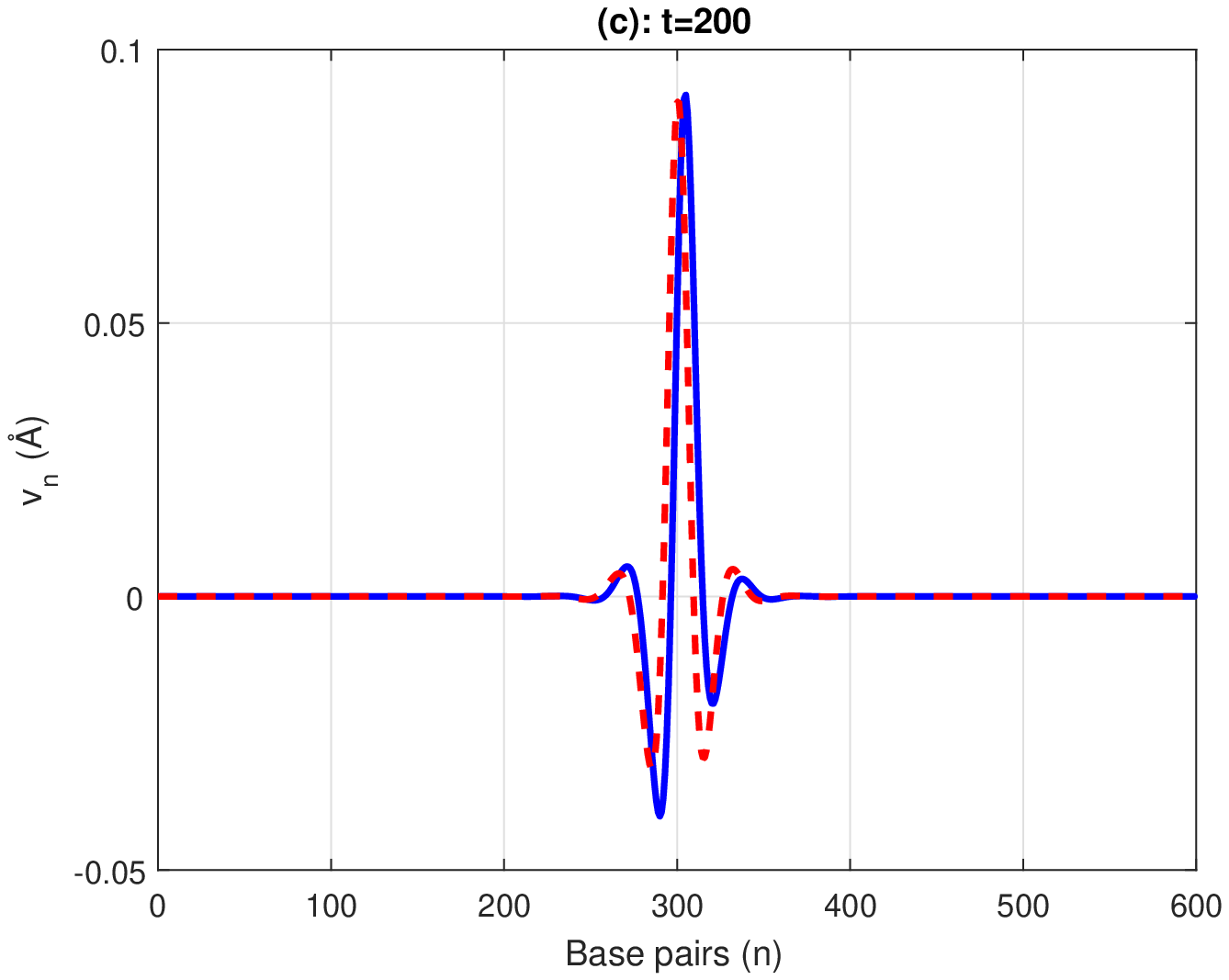}
\includegraphics[width=3in]{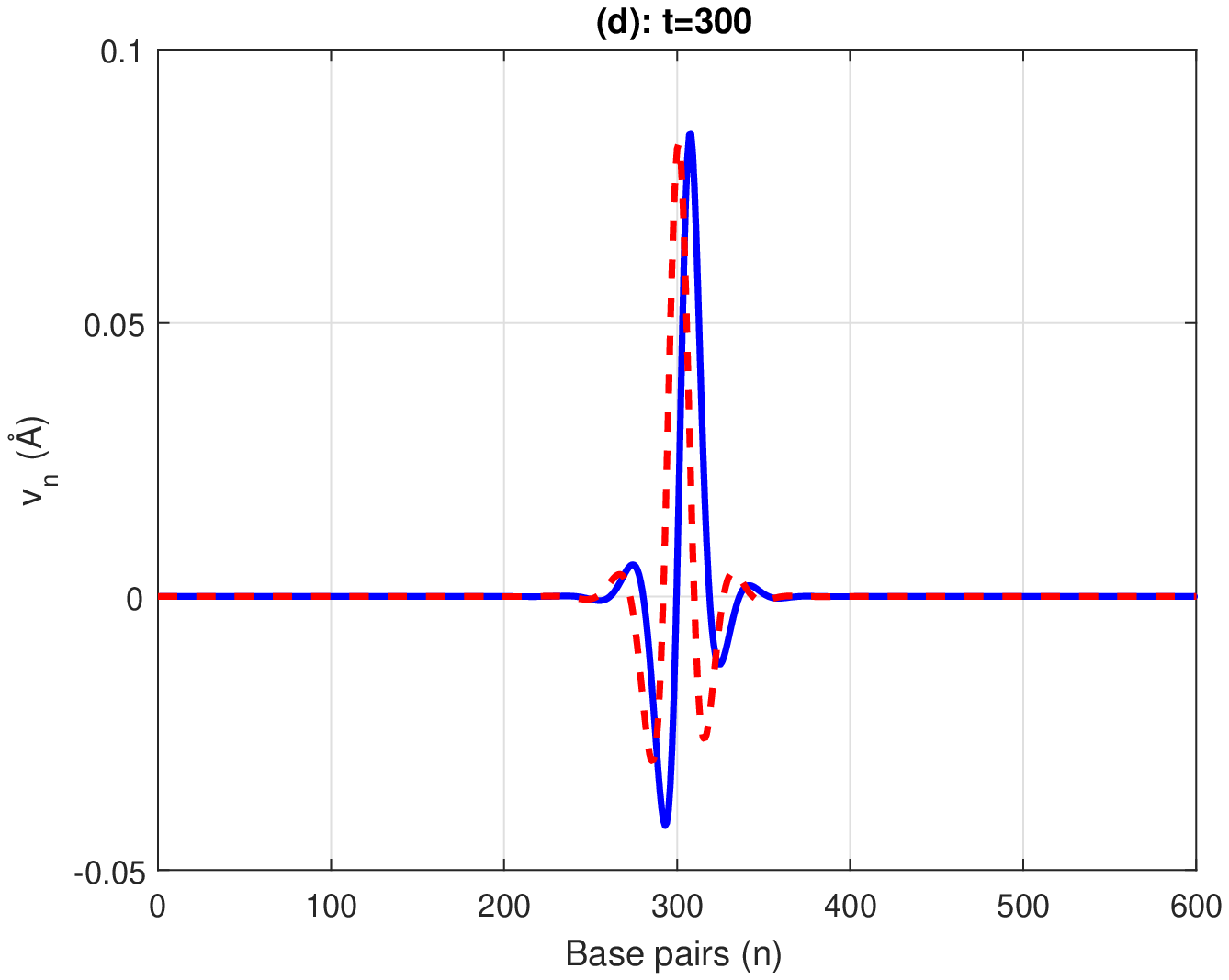}
\caption{(Color online) Comparison between analytical and  numerical solution of the EOM \eqref{eq12} at different time positions for $s=3.00$, $\varepsilon=0.9$, $u_e=1$, $\gamma_0=0.15$ $t.u^{-1}$, $u_c=0.45u_e$ and $q_n^0r=\frac{\pi}{18}$. (solid blue line) the numerical solution. (dash red line) the analytical solution.}\label{fig6}
\end{figure}

The results discussed in the previous section are obtained from the CGL equation (\eqref{eq33}) derived after some approximations and hypothesis and not from the discrete EOM. In order to verify the analytical predictions and check if the above analytical breather soliton can survive in the discrete lattice, the numerical simulations of the discrete EOM (\eqref{eq12}) is carried out by means of a standard fourth-order Runge-Kutta computational scheme with periodic boundary conditions. In our simulations we use the initial condition given by Eq. (\ref{eq46}) and time step $h=2\pi/N$ $t.u.$ with $N=600$. \figref{fig4} presents the $3D$ and $2D$ time-evolution of the solution for a value of LRI parameter $s=3.00$, and we observe the decreasing of the amplitude of the soliton due to the damping forces (\figref{fig4}a). \figref{fig4}b presents the aspect of the numerical solution at $t=115$. As predicted by the analytical results, we observe that the breathing mode appears in DNA molecule when the wave vector is small, namely $q_nr\leq\frac{\pi}{12}$ which corresponds to the domain where the real parts of the angular frequency is different from zero. In \figref{fig5} and \figref{fig6}, we have depicted the $2D$ schematic representation of the analytical and numerical solutions at few time positions for two different wave vectors ($q_nr=\pi/16$ and $q_nr=\pi/18$): $t=0$, $t=50$, $t=200$ and $t=300$. We observe that when the wave vector increases, the width of the soliton grows by increasing the number of base pairs in the bubble. At the same time, the wave amplitude must increase in order to keep the soliton within the lattice length limits. Therefore, small wavenumber leads to a more localized solution. The decreasing of the amplitude of the soliton in time is observed. Also we notice that the shape, the decay of the amplitude and the number of base pairs in the bubble of both solutions after a limited time propagation are the same. But after a long time propagation the number of base pairs which form the bubble remains constant, the shape of the numerical solution is slightly modified also the decay in the amplitude of the numerical solution is less than the theoretical expectations, due to the discreteness effects which usually tend to slow down the motion \cite{dau}. These results confirm that our analytical solution is stable and suitable to predict formation of breather solitons in the DNA molecular chain in the presence of damping and LRI forces.
%
%%%%%%%%%%%%%%%%%%%%%%%%%%%%%%%%%%%%%%%%%%%%%%%%%%%%%%%%%%%%%%%%%%%%%%
%%%%%%%%%%%%%%%%%%%%%%%%%%%%%%%%%%%%%%%%%%%%%%%%%%%%%%%%%%%%%%%%%%%%%%
%
\section{\label{conclusion}Conclusion}
In this work, we have studied the dynamics of breather solitons in a long-range version of the  Peyrard-Bishop DNA model taking into Stokes and hydrodynamical viscous forces. Using the discrete difference operator technique, we have shown that the out-of-phase motion can be described by the CGL equation. As compared to the semi-discrete approach, this technique can lead to the correct expression for the soliton parameters. The breather soliton which represents the opening of base pairs experimentally observed in the DNA molecular chain in the form of bubble, has been found stable when it propagates, however its amplitude decreases due to damping effect. Our numerical simulations have confirmed the validity of the analytical approximate results.
%
%%%%%%%%%%%%%%%%%%%%%%%%%%%%%%%%%%%%%%%%%%%%%%%%%%%%%%%%%%%%%%%%%%%
%%%%%%%%%%%%%%%%%%%%%%%%%%%%%%%%%%%%%%%%%%%%%%%%%%%%%%%%%%%%%%%%%%

%
%Å
%Å
%Å
%Å

\begin{thebibliography}{apssamp}
%{99}
\section*{References}
%
\bibitem{Englander} S.W. Englander \emph{et al}., Proc. Natl. Acad. Sci. U.S.A. \textbf{77}, 7222 (1980).
%
\bibitem{Yomosa1} S. Yomosa, Phys. Rev. A \textbf{27}, 2120 (1983).
%
\bibitem{Homma1} S. Homma  and S. Takeno, Prog. Theor. Phys. \textbf{70}, 308 (1983);.
%
\bibitem{Homma2} S. Takeno and S. Homma,  Prog. Theor. Phys. \textbf{\textbf{72}}, 679 (1984).
%
\bibitem{PB} M. Peyrard and A.R. Bishop, Phys. Rev. Lett. \textbf{62}, 2755 (1989).
%
\bibitem{Dauxois1} T. Dauxois, M. Peyrard, and A.R. Bishop, Phys. Rev. E \textbf{47}, R44 (1993).
%
\bibitem{Dauxois2} T. Dauxois and M. Peyrard, Phys. Rev. E \textbf{51}, 4027 (1995).
%
\bibitem{Campa} A. Campa and A. Giansanti, Phys. Rev. E \textbf{58}, 3585 (1998).
%
\bibitem{dau} T. Dauxois,  Phys. Lett. A. \textbf{l59}, 390 (l991).
%
\bibitem{zdr2} S. Zdravkovi\'{c} and M.V. Satari\'{c}, Chin. Phys. Lett. \textbf{24}, 1210 (2003).
%
\bibitem{zdr22}S. Zdravkovi\'{c} M.V. Satari\'{c} and L. Hadžievski, Chaos \textbf{20}, 043141 (2010).
%
\bibitem{Calla} C. Calladine \emph{et al}., Understanding DNA, 3rd edition (Academic Press, London, 2004).
%
\bibitem{Sha} N.N. Shafranovskaya \emph{et al.}, Pis'ma. Zh. Eksp. Teor. Fiz. \textbf{15}, 404 (1972).
%
\bibitem{rau} D.C. Rau and V.A. Parsegian, Biophysics J. \textbf{61}, 246 (1992).
%
\bibitem{ish} Y. Ishimori, Prog. Theor. Phys. \textbf{68}, 402 (1982).
%
\bibitem{gai1} Yu. B. Gaididei, S.F. Mingaleev, P.L. Christiansen and K. $\emptyset$. Rasmussen, Phys. Lett. A. \textbf{222}, 152 (1996).
%
\bibitem{gai2} S. F. Mingaleev, Yu.B. Gaididei, and F.G. Mertens, Phys. Rev. E. \textbf{58}, 3833 (1998).
%
\bibitem{mvo1} A. Mvogo, G.H. Ben-Bolie and T.C. Kofan\'{e}, Phys. Lett. A. \textbf{378}, 2509 (2014).
%
\bibitem{mvo2} A. Mvogo, G.H. Ben-Bolie and T.C. Kofan\'{e}, Chaos \textbf{25}, 063115 (2015).
%
\bibitem{mil} G. Miloshevich, J.P. Nguenang, T. Dauxois, R. Khomeriki and S. Ruffo, J. Phys. A: Math. Theor. \textbf{50}, 12LT02 (2017).
%
\bibitem{Stenflo} N.R. Pereira and L. Stenflo, Phys. Fluids \textbf{20}, 1733 (1977).
%
\bibitem{fla} S. Flach, Phys. Rev. E \textbf{58}, R4116 (1998).
%
\bibitem{Gai3}  Y.B. Gaididei,  S.F. Mingaleev, P.L. Christiansen  and K.$\emptyset$ Rasmussen, Phys. Rev. E \textbf{55}, 6141 (1997).
%
\bibitem{pey} M. Peyrard, Nonlinearity \textbf{17}, R1 (2004).
%
\bibitem{are} E. Ar\'evalo and F.G. Mertens, Phys. Rev. E. \textbf{67}, 016610 (2003).
%
\bibitem{bru} C. Brunhuber and F.G. Mertens, Phys. Rev. E. \textbf{73}, 016614 (2006).
%
\bibitem{isa} I. Daumont and M. Peyrard, Chaos \textbf{13}, 624 (2003).
%
\bibitem{pey3} M. Peyrard and I. Daumont, Europhys. Lett., \textbf{59} 834 (2002).
%
\bibitem{are2} E. Ar\'evalo Yu. Gaididei and F.G. Mertens, Eur. Phys. J. B \textbf{27}, 63 (2002)
%
\bibitem{boy1} J.W. Boyle, S. A. Nikitov, A.D. Boardman, J.G. Booth and K. Booth, Phys. Rev. B \textbf{53}, 12173 (1996)
%
\bibitem{hoc} L.M. Hocking and K. Stewartson, Proc. R. Soc. London, Ser. A \textbf{326}, 289 (1972).
%
\bibitem{akh} N. Akhmediev and A. Ankiewicz: \emph{Solitons of the Complex Ginzburg-Landau Equation} in Spatial Solitons, Edited by S. Trillo, p. 311 (Springer, New York, 2002).
%
\bibitem{sco} A.C. Scott, F.Y.F. Chu and D.W. McLaughlin, Proc. IEEE. E. \textbf{61}, 1443 (1973).
%
\bibitem{pro1}B.F. Putnam, L.L. Van Zandt, E.W. Prohofsky, K.C. Lu, and W. N. Mei, Biophys. J. \textbf{35}, 271 (1981); E.W. Prohofsky, K.C. Lu, L.L. Van Zandt and B.F. Putnam, Phys. Lett. A \textbf{70}, 492 (1979).
%
%\nocite{*}
\end{thebibliography}
\end{document}